\documentclass[pre,superscriptaddress,longbibliography,aps]{revtex4-2}
\usepackage{graphicx} 
\usepackage{amsmath,amssymb,latexsym,epsfig,graphics,epsf,bm}
\usepackage{graphics,xcolor,tabularx}
\usepackage{float}
\usepackage{booktabs}

\newcommand{\be}{\begin{equation}}
\newcommand{\ee}{\end{equation}}
\newcommand{\fig}[1]{Fig.~\ref{#1}}
\newcommand{\Fig}[1]{Figure~\ref{#1}}
\newcommand{\eq}[1]{Eq.~(\ref{#1})}

\newcommand{\sect}[1]{Sec.~\ref{#1}}
\usepackage[symbol*]{footmisc}

\begin{document} 

\title{Linear-limit aging times of three monoalcohols} 

\author{Jan Philipp Gabriel}
\email{jan.gabriel@dlr.de}
\affiliation{\textit{Glass and Time}, IMFUFA, Department of Science and Environment, Roskilde University, Roskilde, Denmark}
\affiliation{Institute of Materials Physics in Space, German Aerospace Center, 51170 Colone, Germany}
\author{Jeppe C. Dyre}
\email{dyre@ruc.dk}
\affiliation{\textit{Glass and Time}, IMFUFA, Department of Science and Environment, Roskilde University, Roskilde, Denmark}
\author{Tina Hecksher}
\email{tihe@ruc.dk}
\affiliation{\textit{Glass and Time}, IMFUFA, Department of Science and Environment, Roskilde University, Roskilde, Denmark}

\date{\today} 
 
\begin{abstract}
This paper presents data for the physical aging of the three monoalcohols 2-ethyl-1-butanol, 5-methyl-2-hexanol, and 1-phenyl-1-propanol. Aging is studied by monitoring the dielectric loss at a fixed frequency in the kHz range following temperature jumps of a few Kelvin's magnitude, starting from states of equilibrium. The three alcohols differ in the Debye relaxation strength and how much the Debye process is separated from the $\alpha$ process. We first demonstrate that single-parameter aging describes all data well and proceed to utilize this fact to identify the linear-limit normalized aging relaxation functions. From the Laplace transform of these functions, the linear-limit aging loss-peak angular frequency defines the inverse of the linear aging relaxation time. This allows for a comparison to the temperature dependence of the Debye and $\alpha$ dielectric relaxation times of the three monoalcohols. We conclude that the aging response for 5-methyl-2-hexanol and 2-ethyl-1-butanol follows the $\alpha$ relaxation, not the Debye process, while no firm conclusion can be reached for 1-phenyl-1-propanol because its Debye and $\alpha$ processes are too close to be reliably distinguished.
\end{abstract}

\maketitle

\section{Introduction}

Physical aging is a characteristic feature of, in principle, all glasses. It involves the rearrangement of molecules below the glass transition temperature $T_g$, reflecting the fact that any glass continuously approaches the equilibrium liquid state \cite{sim31,kau48}. This can only be observed just below $T_g$, however, because aging is too slow at lower temperatures -- aging of window glass is for instance not an issue. Understanding physical aging to the extent of being able to predict it quantitatively is important for applications of oxide glasses, polymers, metallic glasses, etc., both in production and subsequent use \cite{nar71,maz77,str78,kov79,scherer,hod95,gra12,can13,mic16,mck17,roth,rut17,mauro}.

The physical aging induced by a continuously changing temperature, e.g., as experienced during the production of glasses, is more difficult to model quantitatively than aging following a temperature jump. An ideal temperature jump starts from the glass-forming liquid annealed to equilibrium at the initial temperature, $T_i$, implementing a rapid change of temperature to the final (annealing) temperature, $T_a$, which is kept constant for enough time that the system eventually reaches equilibrium at (we denote the final temperature by $T_a$ and not by $T_f$, which in aging contexts symbolizes the fictive temperature \cite{too46,scherer}). The temperature change should be fast enough that virtually no relaxation takes place before the new temperature by heat conduction is established uniformly throughout the sample. After the jump, aging is monitored by measuring how some physical property, $X=X(t)$, approaches its equilibrium value. In the data reported below the property $X$ is the dielectric loss at a fixed frequency in the kHz range, but $X$ can also be, e.g., the density, the enthalpy, the dc conductivity, an elastic modulus, a linear or nonlinear dielectric property, etc \cite{kov63, nar71, moy76a, str78, che78, ols98, dyr03, dil04, sch91, lun05, bru12, ric15, dil24}. Aging of the structural relaxation time itself can be determined from strain-rate switching experiments, as recently convincingly demonstrated by Ediger and coworkers \cite{ber24}. The only requirement is that accurate measurement of $X$ can be carried out within a short time window during which the system only changes its properties insignificantly.

Achieving close-to-ideal temperature-induced aging conditions is not straightforward, primarily because heat conduction is notoriously slow. We obtain such temperature jumps by working with a thin sample and by quickly changing the temperature by means of a Peltier element placed close to the sample \cite{iga08a}. Recent alternative approaches to studies of physical aging include that of Henot \textit{et al.}, who used Ohmic heating to obtain extremely fast heating rates \cite{henot2023}, and that of Cangialosi and co-workers utilizing the small sample size in fast scanning calorimetry to perform almost instantaneous temperature jumps \cite{dilisio2023,dilisio2024}.

Physical aging is highly nonlinear in the sense that different temperature jumps generally result in quite different normalized relaxation functions, $R(t)$, defined by

\be\label{R_def}
R(t) \,\equiv\, \frac{X(t)-X_a}{X_i-X_a}\,.
\ee
Here $X_i$ and $X_a$ are the equilibrium values of the monitored property at the initial temperature $T_i$ and the final temperature $T_a$, respectively. It is henceforth assumed the jump takes place at $t=0$. Strong nonlinearity is usually observed even for temperature jumps of just 1\% magnitude in the sense that down jumps are much faster and more stretched than up jumps of same magnitude to the same final temperature. This is referred to as the ``asymmetry of approach'', with up jumps being ``autocatalyzed'' and down jumps ``autoretarded'' \cite{kov63,scherer,mau09,can13, mck17}.

The asymmetry of approach has been understood since long as an effect of the so-called material time, which controls the aging and is characterized by a rate of change that itself ages. The rate of aging for a down jump in temperature decreases gradually and converges to the equilibrium aging rate at $T_a$, whereas for an up jump the aging rate increases from a low value to eventually also approach the $T_a$ equilibrium aging rate. A quantitative formalism has existed since 1971 \cite{nar71}, the so-called Tool-Narayanaswamy (TN) formalism, which accounts well for not only the nonlinearity of temperature jumps, but works for any temperature protocol. An excellent account of the TN formalism was given by Scherer in his 1986 textbook \cite{scherer}.

We used above the term ``equilibrium aging rate'', which may seem like a contradiction in terms. Like any other response property, however, physical aging has a linear limit. Thus standard linear-response theory characterized by linearity and time-translational invariance applies for sufficiently small temperature jumps. This limit, which is approached only for temperature changes smaller than about 0.1\%, corresponds to the situation in which the material time is proportional to time throughout the aging process. A fluctuation-dissipation theorem exists also for linear physical aging, meaning that the aging response for very small jumps is determined by the \textit{equilibrium fluctuations} \cite{nie96}. It is in this sense an ``equilibrium aging rate'' may be defined \cite{nis20}.

The linear limit of physical aging was recently reached experimentally by Riechers \textit{et al.} \cite{rie22}, who reported data for aging following temperature jumps of varying amplitude, the smallest being just 10 mK \cite{rie22}. That paper applied a simplified version of the TN formalism, the so-called single-parameter-aging (SPA) formalism from 2015 \cite{hec15}, to predict the results of nonlinear temperature jumps from data for small, linear jumps. Following Ref. \onlinecite{nis20}, we use in this paper SPA to do the opposite: from nonlinear temperature jump data, SPA is utilized to extract the difficult-to-measure linear aging relaxation function.

The investigation presented below involves three monoalcohols. Monoalcohols are generally good glass formers, but they differ from other organic glass-forming liquids by having a Debye process as the slowest and most intense dielectric relaxation process \cite{boh14a}. This is believed to reflect the dynamics of hydrogen-bonded structures, a process that is separate from the ordinary $\alpha$ relaxation associated with structural relaxation \cite{boh14a}. Most other linear-response functions have no -- or only very weak -- relaxations at the dielectric Debye loss-peak frequency; they instead relax on more or less the time scale of the ``ordinary'' $\alpha$ relaxation process \cite{boh14a}. In particular, calorimetric measurements of monoalcohols generally correlate with the $\alpha$ relaxation rather than with the Debye process \cite{huth2007comparing,wang2008calorimetric}.

Since this paper studies the role of the Debye process in aging, the three monoalcohols were selected in order to have different degrees of separation between the Debye and the $\alpha$ relaxation processes as well as different magnitudes of their relative loss magnitudes. In all three cases the OH group is terminal or close to the end of the carbon chain. Two of the alcohols studied are regular simple alcohols, while the third one involves a phenyl group. For further studies it would be interesting to study a system of fixed-length monoalcohols for which the OH group location is varied systematically along the chain.

The question we address below is how physical aging proceeds in monoalcohols: Does aging take place on the time scale of the $\alpha$ process, as one may expect from the general identification of aging as a consequence of structural relaxation \cite{scherer, mck17}, or does aging proceed on the time scale of the slower dielectric Debye process? Since $\alpha$ linear responses may differ in their corresponding relaxation time but otherwise have identical temperature dependence \cite{cutroni2001,jakobsen2005,jakobsen2012communication, roe21}, we focus on comparing the aging relaxation time's \textit{temperature dependence} to those of the Debye and $\alpha$ relaxation times. In order to do this, however, it is necessary to have linear-response aging data at disposal, which are obtained from nonlinear aging data utilizing SPA.

The present work is not the first to address the possible role of the Debye process in physical aging. Previously, the focus was more on the absolute time scales than on their temperature dependence, however, and these studies did not attempt to identify the linear-limit aging time. Already in 2010 Gainaru \textit{et al.} concluded that the dielectric aging of 2-ethyl-1-hexanol follows the $\alpha$ process \cite{gainaru2010coupling}; in contrast, in the non-hydrogen-bonding liquid tri-butyl-phosphate (TBP) with an alcohol-like spectrum created by dielectric cross-correlation, the linear aging time follows the Debye-like mode associated with dielectric cross-correlations \cite{moch2022molecular, gabriel2023comparing}. Reference \onlinecite{pre12} studied mixtures of 2-ethyl-1-hexanol (2E1H) with 2-ethyl-1-hexyl bromide (2E1Br). For an equimolar mixture the Debye and $\alpha$ processes are separated by no less than four decades in frequency, and from a comparison of the involved time scales the authors concluded that physical aging is governed by the $\alpha$ process, not by the Debye process (see also Ref. \onlinecite{gai11}). This is confirmed by the present study on pristine alcohols.

Our investigation proceeds in two steps. First, we validate that the aging data conform to the SPA framework. This has previously been done for glycerol and for several van der Waals bonded liquids, as well as in computer simulations \cite{hec15, nis17, roe19, nis20, meh21, rie22, boh24, hen24}, but not for monoalcohols that as mentioned have quite distinct properties making it highly nontrivial that they should comply to a simple SPA scenario. Once SPA has been validated, we proceed to use it to address whether physical aging in regard to the temperature dependence of the linear-limit aging time follows the Debye process or the $\alpha$ process. Since the purpose is to compare to the temperature dependence of the dielectric Debye and $\alpha$ relaxation times defined by $\tau = 1/(2 \pi \nu_{max})$ in which $\nu_{max}$ is the corresponding loss-peak frequency, we also use this equation to determine the ``equilibrium'' aging relaxation time $\tau_{R}$ from the loss-peak frequency $\nu^R_{max}$ of the frequency response (obtained by Laplace transforming the SPA-extracted linear-limit aging relaxation function).

\section{Methods}

\subsection{Experimental details}

The three monoalcohols studied are: 2-ethyl-1-butanol (``2E1B''; Alfa Aesar, 99 \% purity), 5-methyl-2-hexanol (``5M2H''; Sigma Aldrich, 99 \% purity), and 1-phenyl-1-propanol (``1P1P''; Alfa Aesar, 98+\% purity). All liquids were used as purchased.

We monitored aging after temperature jumps by continuously measuring the dielectric loss by means of a high-resolution Andeen-Hagerling AH2700A bridge, which was done at 1 kHz for 5M2H and at 10 kHz for 2E1B and 1P1P. Thus the dielectric loss is the quantity $X(t)$ in \eq{R_def}. Alternatively, one can use for $X(t)$ the real part of the dielectric constant; this gives somewhat more noisy data but leads otherwise to virtually the same linear aging relaxation times (data not shown). Our aging setup, which allows for accurate temperature control and rapid temperature jumps using a Peltier element for thin samples (50 microns), is described in detail in Ref. \onlinecite{iga08a}. 

The spectra in Figs.~\ref{fig4} and \ref{fig1} were measured with an HP 3488A Multimeter in conjunction with a custom-built frequency generator covering $10^{-2}$-$10^{2}$~Hz and an HP 4284A LCR meter covering $10^{2}$-$10^{6~}$~Hz.

\subsection{Obtaining the linear-limit aging response function from nonlinear temperature jump data via the single-parameter-aging formalism}\label{SPA_sec}

The TN formalism assumes the existence of a material time, $\xi$, with the property that all temperature jumps lead to the same normalized relaxation function $R(t)$ when this function is given in terms of $\xi$. Thus according to TN, if one writes for any specific temperature jump $R(t)=\phi(\xi(t))$, the function $\phi$ is the same for all jumps whereas the functional form of $\xi(t)$ depends (usually strongly) on the jump. If the jump goes from $T_i$ to $T_a=T_i+\Delta T$ where $\Delta T$ is the temperature change, SPA makes the following ansatz for the aging rate $\gamma$ \cite{hec15,hecksher2024}

\be\label{SPA}
\gamma(t)
\,\equiv\,\frac{d\xi}{dt}(t)
\,=\,\gamma_{a}\,e^{-c\,\Delta T R(t)}\,.
\ee
Here $\gamma_{a}$ is the equilibrium linear aging rate at the final temperature of the jump, $T_a$, and $c$ is a material-specific constant. The notation used here is that of Ref.~\onlinecite{hecksher2024}, which differs from Ref. \onlinecite{hec15} that followed the old tradition in the aging field of defining $\Delta T$ as $T_i-T_a$, not as $T_a-T_i$, and which used a dimensionless version of the ``non-linearity parameter'' $c$. Note also that the material time -- and thereby the aging rate $\gamma$ -- is only defined within an arbitrary multiplicative constant.

References \onlinecite{hec15,hec19} showed how to use SPA to calculate the normalized relaxation function of one nonlinear temperature jump from that of another, Ref. \onlinecite{nis20} showed how SPA allows for calculating the linear limit of a nonlinear jump, and Ref. \onlinecite{rie22} showed how SPA can be used for predicting nonlinear jumps from linear ones. Although the reasoning is the same in all cases, we now for completeness briefly summarize how SPA is used for calculating the linear-limit normalized aging relaxation function, $R_{\rm lin}(t)$, from the normalized relaxation function $R(t)$ of a general, nonlinear jump. 

For $R_{\rm lin}(t)$ the expression $R(t)=\phi(\xi(t))$ reduces to $R_{\rm lin}(t_0)=\phi(\gamma_a t_0)$ because $\xi=\gamma_a t_0$ in equilibrium at $T_a$ as well as very close to $T_a$; here and henceforth the laboratory time of a linear-limit temperature jump is denoted by $t_0$ to distinguish it from the laboratory time $t$ of general, nonlinear jumps. Taking the time derivatives leads to $dR/dt=\phi'(\xi(t))\gamma(t)$ and $dR_{\rm lin}/dt_0=\phi'(\xi=\gamma_a t_0)\gamma_a$, respectively. The ratio of these two equations evaluated at times $t$ and $t_0$ corresponding to the same value of $R$ (and thus the same material time $\xi$) is by \eq{SPA} equal to $\exp(-c\Delta T R)$. For the ratio between the time increments $dt$ and $dt_0$ corresponding to the same decrease of $R$, $dR$, one thus has

\be\label{ratio}
\frac{dt_0}{dt}
\,=\,\frac{dR/dt}{dR/dt_0}
\,=\,\frac{\gamma(t)}{\gamma_a}
\,=\,\,e^{-c\,\Delta T R}\,.
\ee
This implies that

\be\label{recipe}
dt
\,=\,e^{c\,\Delta T R}\,dt_0\,.
\ee
For a temperature up jump one has $\Delta T>0$ and the time increment $dt$ needed for a given change of the normalized relaxation function, $dR$, is larger than the time increment $dt_0$ needed for the same change of the linear relaxation function at $T_a$. The opposite applies for a down jump. In both cases \eq{recipe} provides the SPA recipe for calculating $R_{\rm lin}(t)$ from nonlinear jump data in the form of $R(t)$. This is done in a step-by-step manner starting at $t=0$ by transforming the time axis of $R(t)$ to that of $R_{\rm lin}(t)$ by means of \eq{recipe} \cite{hec15,roe19,rie22}. Since $R_{\rm lin}(t)$ for all jumps to $T_a$ must be the same function, the material-characteristic nonlinearity parameter $c$ can be determined by optimizing the collapse of the $R_{\rm lin}(t)$ functions for jumps to the same temperature. This is done for all final temperatures $T_a$; $c$ is then the nonlinearity parameter that results in the best overall collapse.

\section{Results and discussion}

\begin{figure*}[t!]
	\centering
	\includegraphics[width=17cm]{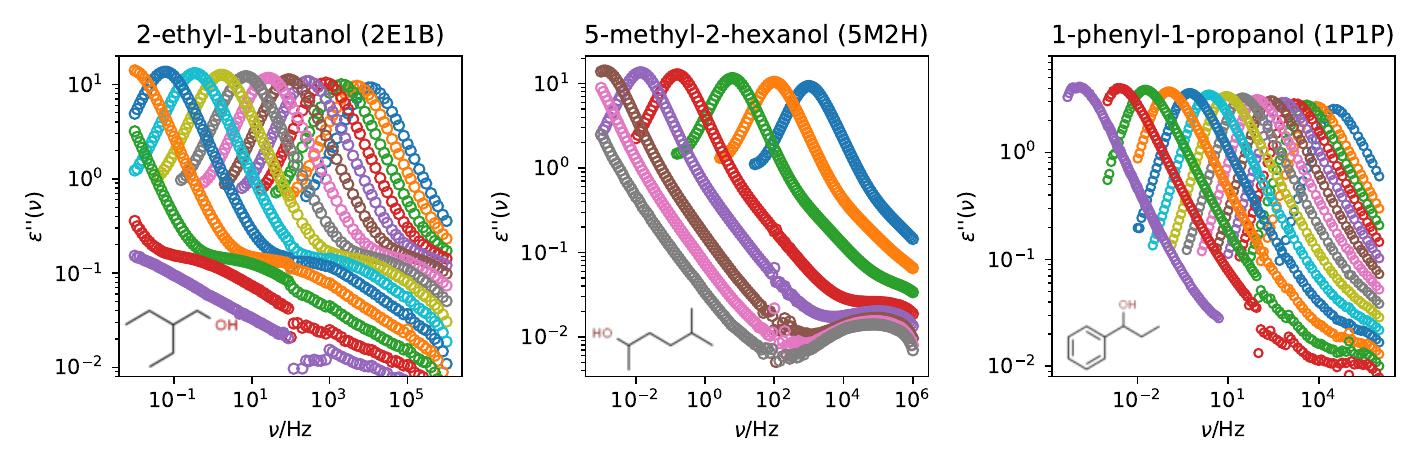}
	\caption{Equilibrium dielectric spectra at different temperatures of 2E1B, 5M2H, and 1P1P. The figure shows log-log plots of the imaginary part of the frequency-dependent dielectric constant, $\varepsilon''$, as a function of the frequency $\nu$. The insets show the molecular structures of each monoalcohol.}\label{fig4}
\end{figure*}

\begin{table}[H]\centering
\scriptsize
\begin{tabular}{@{}c|c|c@{}}
\textbf{2E1B {[}K{]}} & \textbf{5M2H {[}K{]}} & \textbf{1P1P {[}K{]}} \\ \midrule
200                   & 193                   & 230                   \\
195                   & 183                   & 225                   \\
190                   & 173                   & 222.5                 \\
185                   & 163                   & 220                   \\
180                   & 158                   & 217.5                 \\
175                   & 154                   & 215                   \\
170                   & 152                   & 212.5                 \\
165                   & 150                   & 210                   \\
160                   &                       & 207.5                 \\
155                   &                       & 205                   \\
150                   &                       & 202.5                 \\
145                   &                       & 200                   \\
140                   &                       & 197.5                 \\
135                   &                       & 195                   \\
130                   &                       & 193               
\end{tabular}
\caption{Temperatures for the equilibrium dielectric spectra of \fig{fig4}.}
\label{tab1}
\end{table}

\begin{figure}[H]
	\centering
	\includegraphics[width=8.4cm]{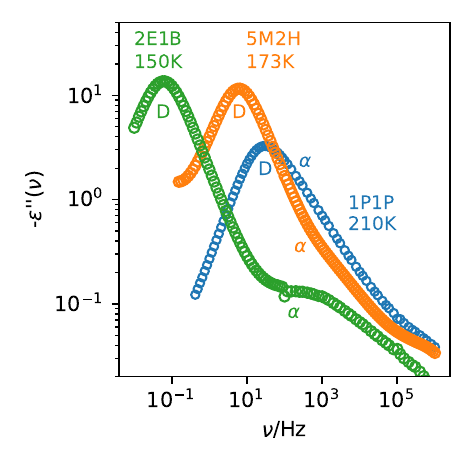}
	\caption{Examples of dielectric loss peaks of 2E1B (green), 5M2H (orange), and 1P1P (blue). The Debye peak is marked by the letter D and the $\alpha$ peak by an $\alpha$. For 2E1B the latter is clearly visible, while it for 5M2H is manifested as a change of slope. In contrast, the 1P1P loss peak is asymmetric and looks much like those of typical non-monoalcohol organic glass formers \cite{nie09,pab21}. For 1P1P supplementary dynamic light scattering measurements have revealed the existence of separate Debye and $\alpha$ peaks, albeit close to each other \cite{boh14a,hansen1997dynamics}.}
	\label{fig1}
\end{figure}

\Fig{fig4} shows equilibrium dielectric loss spectra of the three monoalcohols for a range of temperatures. To compare the liquids we plot in \fig{fig1} three representative loss spectra. 2E1B and 5M2H have large, symmetric, narrow loss peaks. These are the noted Debye loss peaks (marked by D), the above-discussed characteristic feature of monoalcohols \cite{boh14a}. For 2E1B there is clearly an additional high-frequency process, which is identified as the $\alpha$ process \cite{boh14a}. An underlying $\alpha$ process is also visible in the 5M2H data, though it is here much less pronounced and merely visible as a change of slope. The 1P1P loss peak, on the other hand, has no clearly visible Debye process. Its peak is broader and asymmetric to the high-frequency side with a high-frequency slope that is not far from -0.5 \cite{nie09,pab21}; thus the 1P1P loss peak looks like typical peaks of van der Waals bonded glass-forming liquids \cite{nie09,ric15}. If there is a dielectric Debye process in 1P1P, it has merged completely with the $\alpha$ process. However, a closer analysis involving also light-scattering data, which are known to mainly probe the self part of the dipole time-autocorrelation function \cite{pab21}, reveals that the dielectric spectrum of 1P1P may also be perceived as a sum of a Debye and an $\alpha$ peak \cite{boh14a,hansen1997dynamics}.

\begin{figure}[H]
	\centering
	\includegraphics[width=0.9\textwidth]{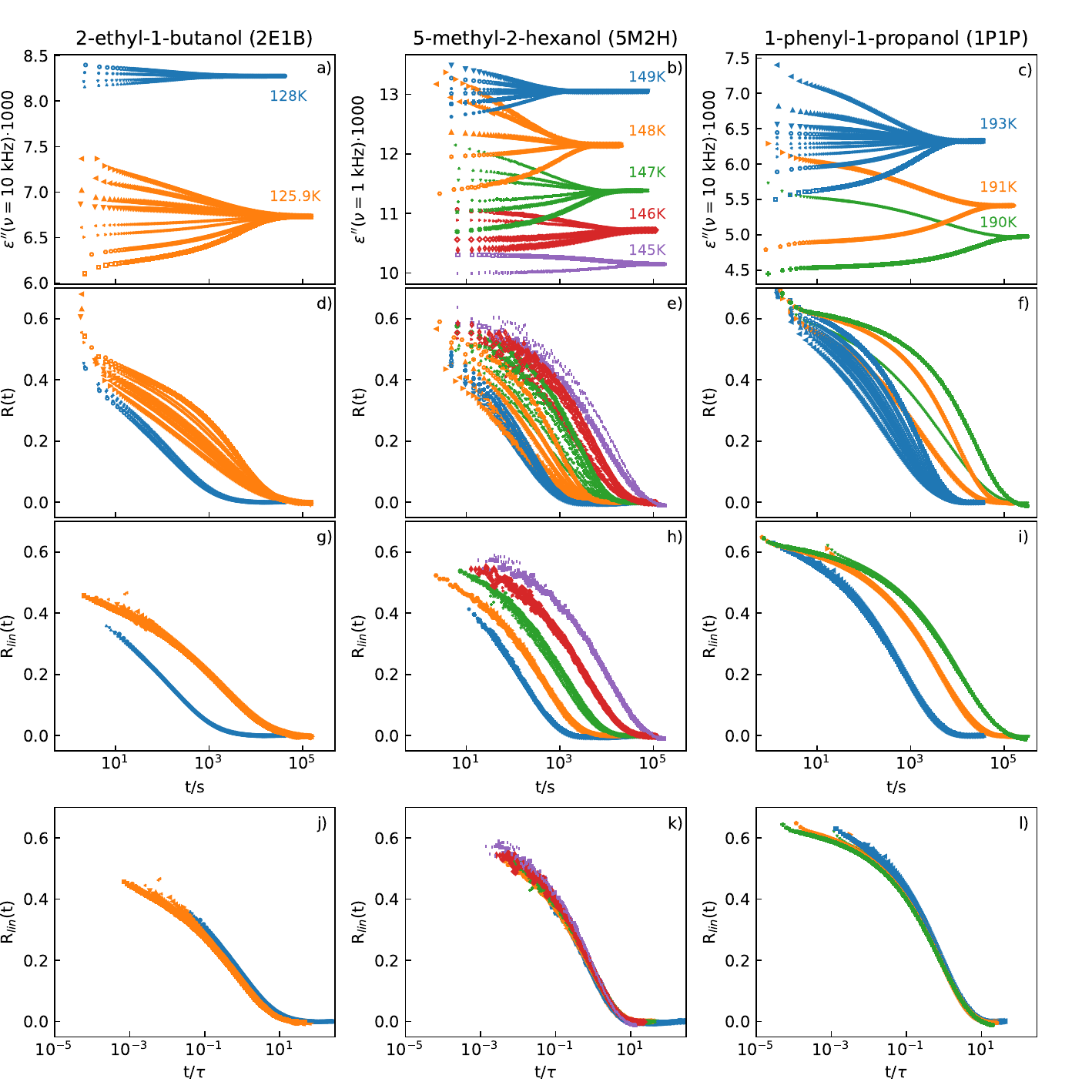}
	\caption{Aging data and SPA analysis. The left panels show data for 2E1B, the middle panels for 5M2H, the right panels for 1P1P. (a)-(c) show data for the imaginary part of the dielectric susceptibility measured at the indicated frequency after a temperature jump at $t=0$. 
    (d)-(f) show the normalized relaxation functions corresponding to the data of (a)-(c) defined by \eq{R_def}.
    (g)-(i) show for each final temperature collapse of the different jump data when these are transformed into a linear aging relaxation function by means of \eq{recipe}. This transformation involves the material-specific parameter, $c$, which for each liquid is determined to optimize the data collapse for jumps to the same temperature. 
    (j)-(l) collapse the linear aging relaxation function for different target temperatures onto a single one by empirically rescaling the time axis. The data collapse seen here confirms time-temperature superposition, a prerequisite of the TN formalism and therefore also for SPA. }	\label{fig2}
\end{figure}

\begin{table}[H]\centering
	\scriptsize
	\begin{tabular}{ccc|ccc|ccc}
		\multicolumn{3}{c|}{\textbf{2E1B}}         & \multicolumn{3}{c|}{\textbf{5M2H}}       & \multicolumn{3}{c}{\textbf{1P1P}}        \\ \hline
		T$_{i}$/K & T$_{a}$/K & $\Delta$T/K & T$_{i}$/K & T$_{a}$/K & $\Delta$T/K & T$_{i}$/K & T$_{a}$/K & $\Delta$T/K \\ \hline
		127.8        & 128          & 0.2        & 148.9        & 149        & 0.1        & 192.9        & 193        & 0.1        \\
		127.5        & 128          & 0.5        & 148.5        & 149        & 0.5        & 192.75       & 193        & 0.25        \\
		125.4        & 125.9        & 0.5        & 148          & 149        & 1.0        & 192.5        & 193        & 0.5        \\
		125          & 125.9        & 0.9        & 147.5        & 148        & 0.5        & 192          & 193        & 1.0        \\
		124.1        & 125.9        & 1.8        & 146.1        & 148        & 1.9        & 191          & 193        & 2.0        \\
		123.2        & 125.9        & 2.7        & 145.1        & 148        & 2.9        & 189          & 191        & 2.0        \\
		127.8        & 128          & 0.2        & 146.5        & 147        & 0.5        & 188          & 190        & 2.0        \\
		128.5        & 128          & -0.5         & 146          & 147      & 1.0        & 193.1        & 193        & -0.1         \\
		126          & 125.9        & -0.1         & 145.1        & 147      & 1.9        & 193.25       & 193        & -0.25         \\
		126.4        & 125.9        & -0.5         & 145.5        & 146      & 0.5        & 193.5        & 193        & -0.5         \\
		126.8        & 125.9        & -0.9         & 145          & 146      & 1.0        & 194          & 193        & -1.0         \\
		127.8        & 125.9        & -1.9         & 144.5        & 145      & 0.5        & 195          & 193        & -2.0         \\
		&            &              & 149.1        & 149        & -0.1         & 193          & 191        & 2.0         \\
		&            &              & 149.5        & 149        & -0.5         & 192          & 190        & 2.0         \\
		&            &              & 150          & 149        & -1.0         &              &            &              \\
		&            &              & 148.5        & 148        & -0.5         &              &            &              \\
		&            &              & 149.9        & 148        & -1.9         &              &            &              \\
		&            &              & 147.5        & 147        & -0.5         &              &            &              \\
		&            &              & 148          & 147        & -1.0         &              &            &              \\
		&            &              & 148.9        & 147        & -1.9         &              &            &              \\
		&            &              & 146.5        & 146        & -0.5         &              &            &              \\
		&            &              & 147          & 146        & -1.0         &              &            &              \\
		&            &              & 145.5        & 145        & -0.5         &              &            &             
    \end{tabular}
	\caption{List of all temperature jumps performed, from the initial temperature T$_i$ to the final temperature T$_a$. The jump magnitudes $\Delta T\equiv T_a-T_i$ are also listed.}
    \label{tab2}
\end{table}

The aging data and their analysis are presented in \fig{fig2} devoting one column to each liquid. The top three panels show all temperature jump data, i.e., the imaginary parts of the dielectric permittivity measured at a fixed frequency in the kHz region plotted as a function of the logarithm of the time since the jump was initiated. For 2E1B there are jumps to two final temperatures, for 5M2H to five final temperatures, and for 1P1P to three final temperatures. The jump magnitudes vary from 0.1 K to 2.9 K (Table \ref{tab2}). 

Using $X=\varepsilon''(\nu_0,t)$ in which $\nu_0=1$~kHz or 10~kHz as the monitored variable, the second row of \fig{fig2} shows all the normalized relaxation functions $R(t)$, defined by \eq{R_def}, plotted as functions of time. Note that the normalized relaxation functions do not start in unity at short times; this is because there is a rapid change of the dielectric loss that is too fast to be probed by the setup. This rapid change is notably larger for 2E1B than for the two other liquids. We interpret the rapid jump as reflecting the existence of a fast structural relaxation process (an aging ``$\beta$ process''), which apparently is largest for 2E1B. We have no good explanation for this difference, which it could be interesting to investigate further, e.g., by comparing to the magnitude of possible dielectric $\beta$ processes.

The functions $R(t)$ do not collapse, even for jumps to the same temperature (same color), which illustrates the above-mentioned strong nonlinearity of physical aging \cite{scherer}. In order to arrive at the linear-limit normalized relaxation function, $R_{\rm lin}(t)$, we employ single-parameter aging (SPA) as detailed in \sect{SPA_sec}. The fitting procedure involves one free parameter for each liquid, the ``nonlinearity parameter'' $c$ in \eq{SPA}. SPA uniquely determines $R_{\rm lin}(t)$ from the normalized relaxation function of any jump, $R(t)$. For each liquid, $c$ is found by searching for the best collapse of the calculated linear-limit aging relaxation functions for jumps to the same temperature. This results in $c=1.72\, \textrm{K}^{-1}$ for 2E1B and $c=1.19\,\textrm{K}^{-1}$ for both 5M2H and 1P1P. 

For these $c$ values the linear-limit normalized relaxation functions, $R_{\rm lin}(t)$, are shown in the third row of \fig{fig2}. There is a good data collapse, which validates SPA. In the fourth row we empirically scale the time axis, which for all three liquids results in a good overall data collapse. This provides a consistency check of the analysis, because the TN formalism -- and thus SPA -- assumes time-temperature superposition for the linear normalized relaxation functions. 

\begin{figure}[H]
    \centering
    \includegraphics[width=17cm]{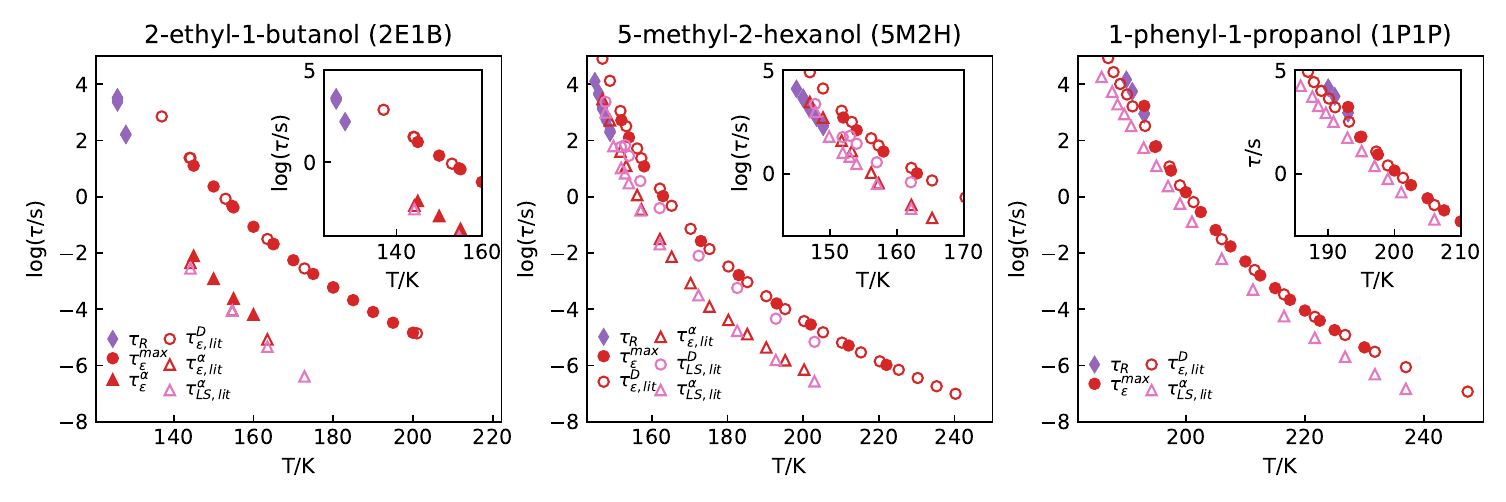}
    \caption{Characteristic times $\tau$ plotted as functions of temperature for the three monoalcohols, including here also light-scattering data. The full symbols are defined as follows (compare the lower left corners):
    $\tau_R$ is the linear-limit aging time calculated as one over the loss-peak angular frequency of the Laplace transform of $R_{\rm lin}(t)$ determined from the nonlinear aging data by SPA via \eq{SPA} (though only visible upon magnification, the figure reports one point for each temperature jump);
    $\tau^{max}_\varepsilon$ is the inverse dielectric loss-peak angular frequency;
    $\tau^{\alpha}_\varepsilon$ is the dielectric $\alpha$ relaxation time estimated as the inverse dielectric loss-peak angular frequency (identified by visual inspection); the open symbols are data from the literature \cite{bauer2013debye,Gabriel:2018a,Gabriel:2018,Gabriel:2018c,Gabriel:2018a}.
    For 2E1B and 5M2H the linear aging times follow the $\alpha$ times, not the Debye times, in their values and temperature dependencies.
    For 1P1P, which is characterized by an almost complete merging of the dielectric Debye and $\alpha$ processes (\fig{fig1}), it would be premature to conclude anything about the temperature dependence. We note, however, that the linear-limit aging times are closer to the Debye times than to the $\alpha$ times.    }
    \label{fig3}
\end{figure}

From the functions $R_{\rm lin}(t)$ we extract the linear-limit aging relaxation time at each final temperature as the inverse of the loss-peak angular frequency of the Laplace transform of $R_{\rm lin}(t)$. Having done so, we can now address this paper's scientific question: Does the temperature dependence of the linear aging time, $\tau_R$, follow that of the $\alpha$ process -- as most other physical processes do \cite{boh14a,jakobsen2012communication,roe21} -- or does it follow the slower Debye process \cite{gainaru2010coupling,moch2022molecular, gabriel2023comparing}? To answer this we plot for each liquid in \fig{fig3} the following quantities as functions of temperature: the Debye relaxation times \cite{bauer2013debye,Gabriel:2018a,Gabriel:2018,Gabriel:2018c}, the $\alpha$ process relaxation times extracted by dielectric spectroscopy \cite{bauer2013debye,Gabriel:2018a,Gabriel:2018,Gabriel:2018c}, and the dynamic-light-scattering relaxation times \cite{Gabriel:2018a,Gabriel:2018,Gabriel:2018c}.

In general, different linear-response functions may have different characteristic relaxation times \cite{cutroni2001,jakobsen2005,jakobsen2012communication, roe21}, i.e., the same underlying process can have different relaxation times when observed with different techniques. A trivial example of this is the comparison of the compliance and modulus representation of the same data: depending on the relaxation strength, this can lead to quite different relaxation times. For several liquids, a time-scale ordering has been established in which the thermal response functions, i.e., response functions where the input is a temperature modulation (dynamic heat capacity, thermal expansion coefficient, etc) -- are generally slower than the dielectric response functions by up to one decade \cite{jakobsen2012communication, roe21}. This means that it is not possible to reach a conclusion on this paper's scientific question by comparing time scales alone. Thus in order to determine whether physical aging follows the Debye or the $\alpha$ process, we must take into account not primarily the proximity of the relaxation times, but how the relaxation times vary with temperature.

The aging data cover a smaller temperature range than the other data of \fig{fig3}. The figure nevertheless illuminates the physics of aging and its relation to dielectric processes. For the two monoalcohols with the most pronounced and clearly separated Debye processes, 2E1B and 5M2H, the conclusion is clear: Physical aging follows the $\alpha$ process in its temperature dependence, not the Debye process. For 2E1B there is no overlap between the $\alpha$ relaxation times extracted from the dielectric spectra and the aging relaxation time, but the separation between the Debye time scale and the aging relaxation time clearly exceeds 2-3 decades and the aging time follows the temperature dependence of the dielectric $\alpha$ relaxation time rather than of the dielectric Debye relaxation time. In the case of 5M2H, the $\alpha$ relaxation time found by dielectrics and light scattering is identical to the linear aging relaxation time: In this case both time scale and temperature dependence match. For 1P1P the picture is less clear. Here separate $\alpha$ and Debye processes have been determined from data for non-linear dielectric spectroscopy \cite{gabriel2021high} and light scattering \cite{boehmer2019influence}. These data indicate that aging follows the Debye and not the $\alpha$ process in the relaxation times, but we hesitate to conclude that this reflects a causal relation because the $\alpha$ and the Debye dielectric processes of 1P1P are so close that the time-scale determination is associated with sizable uncertainty. In particular, the temperature dependence of the two processes cannot be reliably distinguished. 

The fact that the Debye process in 2E1B and 5M2H appears to have no relation to the linear aging time scale while the situation possibly is different for 1P1P, suggests that different origins of the cross-correlations responsible for the Debye process are at play. In other words, our results indicate that there different classes of Debye processes may exist: In some systems the linear aging time scale clearly follows the dielectric $\alpha$ relaxation, in other systems it is closer to the dielectric Debye process. Such a difference makes sense in view of the fact that hydrogen structures evolve very differently in different systems in the form of chains \cite{Dannhauser1968a}, rings \cite{Dannhauser1968a,Singh2012}, or in a branched \cite{Sillren2012} form. Thus 2E1B and 2E1H are classical chain-like alcohols, while 1P1P is believed to from a mixture of ring- and chain-like structures. Note also that additional cross-correlations can form in phenyl alcohols due to the pi-pi electron systems \cite{nowok2021influence,nowok2021molecular}, as has also been observed in recent simulations \cite{grelska2023computer}.

\section{Summary}

We have presented data for the physical aging of three monoalcohols obtained by monitoring the high-frequency dielectric loss after up and down temperature jumps of small and moderate magnitude. The aging of all three liquids conforms to SPA. This allows to extract the linear-limit normalized aging relaxation functions $R_{\rm lin}(t)$ at each final temperature. From these functions, we determined the linear-limit aging relaxation times as the inverse of the loss-peak angular frequencies of the Laplace transform of $R_{\rm lin}(t)$. Comparing to the temperature dependence of the Debye and $\alpha$ relaxation inverse loss peak frequencies, the analysis showed that for the two liquids with best separation of the Debye and the $\alpha$ processes, 2E1B and 5M2H, physical aging follows the $\alpha$ process and not the Debye process. The third liquid, 1P1P, appears to follow the Debye process in its aging-time temperature dependence, but in this liquid no definite conclusion can be made because the Debye and $\alpha$ relaxation times are quite close.

\begin{acknowledgments}
This work was supported by the VILLUM Foundation's \textit{Matter} grant (VIL16515).
\end{acknowledgments}


\begin{thebibliography}{71}%
	\makeatletter
	\providecommand \@ifxundefined [1]{%
		\@ifx{#1\undefined}
	}%
	\providecommand \@ifnum [1]{%
		\ifnum #1\expandafter \@firstoftwo
		\else \expandafter \@secondoftwo
		\fi
	}%
	\providecommand \@ifx [1]{%
		\ifx #1\expandafter \@firstoftwo
		\else \expandafter \@secondoftwo
		\fi
	}%
	\providecommand \natexlab [1]{#1}%
	\providecommand \enquote  [1]{``#1''}%
	\providecommand \bibnamefont  [1]{#1}%
	\providecommand \bibfnamefont [1]{#1}%
	\providecommand \citenamefont [1]{#1}%
	\providecommand \href@noop [0]{\@secondoftwo}%
	\providecommand \href [0]{\begingroup \@sanitize@url \@href}%
	\providecommand \@href[1]{\@@startlink{#1}\@@href}%
	\providecommand \@@href[1]{\endgroup#1\@@endlink}%
	\providecommand \@sanitize@url [0]{\catcode `\\12\catcode `\$12\catcode
		`\&12\catcode `\#12\catcode `\^12\catcode `\_12\catcode `\%12\relax}%
	\providecommand \@@startlink[1]{}%
	\providecommand \@@endlink[0]{}%
	\providecommand \url  [0]{\begingroup\@sanitize@url \@url }%
	\providecommand \@url [1]{\endgroup\@href {#1}{\urlprefix }}%
	\providecommand \urlprefix  [0]{URL }%
	\providecommand \Eprint [0]{\href }%
	\providecommand \doibase [0]{https://doi.org/}%
	\providecommand \selectlanguage [0]{\@gobble}%
	\providecommand \bibinfo  [0]{\@secondoftwo}%
	\providecommand \bibfield  [0]{\@secondoftwo}%
	\providecommand \translation [1]{[#1]}%
	\providecommand \BibitemOpen [0]{}%
	\providecommand \bibitemStop [0]{}%
	\providecommand \bibitemNoStop [0]{.\EOS\space}%
	\providecommand \EOS [0]{\spacefactor3000\relax}%
	\providecommand \BibitemShut  [1]{\csname bibitem#1\endcsname}%
	\let\auto@bib@innerbib\@empty
	\bibitem [{\citenamefont {Simon}(1931)}]{sim31}%
	\BibitemOpen
	\bibfield  {author} {\bibinfo {author} {\bibfnamefont {F.}~\bibnamefont
			{Simon}},\ }\bibfield  {title} {\bibinfo {title} {{{\"U}ber den Zustand der
				unterk{\"u}hlten Fl{\"u}ssigkeiten und Gl{\"a}ser}},\ }\href@noop {}
	{\bibfield  {journal} {\bibinfo  {journal} {Z. Anorg. Allg. Chem.}\ }\textbf
		{\bibinfo {volume} {203}},\ \bibinfo {pages} {219} (\bibinfo {year}
		{1931})}\BibitemShut {NoStop}%
	\bibitem [{\citenamefont {Kauzmann}(1948)}]{kau48}%
	\BibitemOpen
	\bibfield  {author} {\bibinfo {author} {\bibfnamefont {W.}~\bibnamefont
			{Kauzmann}},\ }\bibfield  {title} {\bibinfo {title} {The nature of the glassy
			state and the behavior of liquids at low temperatures},\ }\href
	{https://doi.org/10.1021/cr60135a002} {\bibfield  {journal} {\bibinfo
			{journal} {Chem. Rev.}\ }\textbf {\bibinfo {volume} {43}},\ \bibinfo {pages}
		{219} (\bibinfo {year} {1948})}\BibitemShut {NoStop}%
	\bibitem [{\citenamefont {Narayanaswamy}(1971)}]{nar71}%
	\BibitemOpen
	\bibfield  {author} {\bibinfo {author} {\bibfnamefont {O.~S.}\ \bibnamefont
			{Narayanaswamy}},\ }\bibfield  {title} {\bibinfo {title} {A model of
			structural relaxation in glass},\ }\href
	{https://doi.org/10.1111/j.1151-2916.1971.tb12186.x} {\bibfield  {journal}
		{\bibinfo  {journal} {J. Amer. Ceram. Soc.}\ }\textbf {\bibinfo {volume}
			{54}},\ \bibinfo {pages} {491} (\bibinfo {year} {1971})}\BibitemShut
	{NoStop}%
	\bibitem [{\citenamefont {{Mazurin}}(1977)}]{maz77}%
	\BibitemOpen
	\bibfield  {author} {\bibinfo {author} {\bibfnamefont {O.}~\bibnamefont
			{{Mazurin}}},\ }\bibfield  {title} {\bibinfo {title} {{Relaxation phenomena
				in glass}},\ }\href {https://doi.org/10.1016/0022-3093(77)90092-8} {\bibfield
		{journal} {\bibinfo  {journal} {J. Non-Cryst. Solids}\ }\textbf {\bibinfo
			{volume} {25}},\ \bibinfo {pages} {129} (\bibinfo {year} {1977})}\BibitemShut
	{NoStop}%
	\bibitem [{\citenamefont {Struik}(1978)}]{str78}%
	\BibitemOpen
	\bibfield  {author} {\bibinfo {author} {\bibfnamefont {L.~C.~E.}\
			\bibnamefont {Struik}},\ }\href@noop {} {\emph {\bibinfo {title} {{Physical
					Aging in Amorphous Polymers and Other Materials}}}}\ (\bibinfo  {publisher}
	{Elsevier, Amsterdam},\ \bibinfo {year} {1978})\BibitemShut {NoStop}%
	\bibitem [{\citenamefont {Kovacs}\ \emph {et~al.}(1979)\citenamefont {Kovacs},
		\citenamefont {Aklonis}, \citenamefont {Hutchinson},\ and\ \citenamefont
		{Ramos}}]{kov79}%
	\BibitemOpen
	\bibfield  {author} {\bibinfo {author} {\bibfnamefont {A.~J.}\ \bibnamefont
			{Kovacs}}, \bibinfo {author} {\bibfnamefont {J.~J.}\ \bibnamefont {Aklonis}},
		\bibinfo {author} {\bibfnamefont {J.~M.}\ \bibnamefont {Hutchinson}},\ and\
		\bibinfo {author} {\bibfnamefont {A.~R.}\ \bibnamefont {Ramos}},\ }\bibfield
	{title} {\bibinfo {title} {Isobaric volume and enthalpy recovery of glasses.
			{II. A} transparent multiparameter theory},\ }\href
	{https://doi.org/10.1002/pol.1979.180170701} {\bibfield  {journal} {\bibinfo
			{journal} {J. Polym. Sci. Polym. Phys.}\ }\textbf {\bibinfo {volume} {17}},\
		\bibinfo {pages} {1097} (\bibinfo {year} {1979})}\BibitemShut {NoStop}%
	\bibitem [{\citenamefont {Scherer}(1986)}]{scherer}%
	\BibitemOpen
	\bibfield  {author} {\bibinfo {author} {\bibfnamefont {G.~W.}\ \bibnamefont
			{Scherer}},\ }\href@noop {} {\emph {\bibinfo {title} {{Relaxation in Glass
					and Composites}}}}\ (\bibinfo  {publisher} {Wiley, New York},\ \bibinfo
	{year} {1986})\BibitemShut {NoStop}%
	\bibitem [{\citenamefont {Hodge}(1995)}]{hod95}%
	\BibitemOpen
	\bibfield  {author} {\bibinfo {author} {\bibfnamefont {I.~M.}\ \bibnamefont
			{Hodge}},\ }\bibfield  {title} {\bibinfo {title} {Physical aging in polymer
			glasses},\ }\href {https://doi.org/10.1126/science.267.5206.1945} {\bibfield
		{journal} {\bibinfo  {journal} {Science}\ }\textbf {\bibinfo {volume}
			{267}},\ \bibinfo {pages} {1945} (\bibinfo {year} {1995})}\BibitemShut
	{NoStop}%
	\bibitem [{\citenamefont {Grassia}\ and\ \citenamefont {Simon}(2012)}]{gra12}%
	\BibitemOpen
	\bibfield  {author} {\bibinfo {author} {\bibfnamefont {L.}~\bibnamefont
			{Grassia}}\ and\ \bibinfo {author} {\bibfnamefont {S.~L.}\ \bibnamefont
			{Simon}},\ }\bibfield  {title} {\bibinfo {title} {{Modeling volume relaxation
				of amorphous polymers: Modification of the equation for the relaxation time
				in the KAHR model}},\ }\href
	{https://doi.org/{10.1016/j.polymer.2012.06.013}} {\bibfield  {journal}
		{\bibinfo  {journal} {Polymer}\ }\textbf {\bibinfo {volume} {53}},\ \bibinfo
		{pages} {3613} (\bibinfo {year} {{2012}})}\BibitemShut {NoStop}%
	\bibitem [{\citenamefont {Cangialosi}\ \emph {et~al.}(2013)\citenamefont
		{Cangialosi}, \citenamefont {Boucher}, \citenamefont {Alegria},\ and\
		\citenamefont {Colmenero}}]{can13}%
	\BibitemOpen
	\bibfield  {author} {\bibinfo {author} {\bibfnamefont {D.}~\bibnamefont
			{Cangialosi}}, \bibinfo {author} {\bibfnamefont {V.~M.}\ \bibnamefont
			{Boucher}}, \bibinfo {author} {\bibfnamefont {A.}~\bibnamefont {Alegria}},\
		and\ \bibinfo {author} {\bibfnamefont {J.}~\bibnamefont {Colmenero}},\
	}\bibfield  {title} {\bibinfo {title} {{Physical aging in polymers and
				polymer nanocomposites: recent results and open questions}},\ }\href
	{https://doi.org/10.1039/C3SM51077H} {\bibfield  {journal} {\bibinfo
			{journal} {Soft Matter}\ }\textbf {\bibinfo {volume} {9}},\ \bibinfo {pages}
		{8619} (\bibinfo {year} {2013})}\BibitemShut {NoStop}%
	\bibitem [{\citenamefont {Micoulaut}(2016)}]{mic16}%
	\BibitemOpen
	\bibfield  {author} {\bibinfo {author} {\bibfnamefont {M.}~\bibnamefont
			{Micoulaut}},\ }\bibfield  {title} {\bibinfo {title} {Relaxation and physical
			aging in network glasses: a review},\ }\href
	{https://doi.org/10.1088/0034-4885/79/6/066504} {\bibfield  {journal}
		{\bibinfo  {journal} {Rep. Prog. Phys.}\ }\textbf {\bibinfo {volume} {79}},\
		\bibinfo {pages} {066504} (\bibinfo {year} {2016})}\BibitemShut {NoStop}%
	\bibitem [{\citenamefont {McKenna}\ and\ \citenamefont {Simon}(2017)}]{mck17}%
	\BibitemOpen
	\bibfield  {author} {\bibinfo {author} {\bibfnamefont {G.~B.}\ \bibnamefont
			{McKenna}}\ and\ \bibinfo {author} {\bibfnamefont {S.~L.}\ \bibnamefont
			{Simon}},\ }\bibfield  {title} {\bibinfo {title} {50th anniversary
			perspective: {Challenges} in the dynamics and kinetics of glass-forming
			polymers},\ }\href {https://doi.org/10.1021/acs.macromol.7b01014} {\bibfield
		{journal} {\bibinfo  {journal} {Macromolecules}\ }\textbf {\bibinfo {volume}
			{50}},\ \bibinfo {pages} {6333} (\bibinfo {year} {2017})}\BibitemShut
	{NoStop}%
	\bibitem [{\citenamefont {Roth}(2017)}]{roth}%
	\BibitemOpen
	\bibinfo {editor} {\bibfnamefont {C.~B.}\ \bibnamefont {Roth}},\ ed.,\
	\href@noop {} {\emph {\bibinfo {title} {Polymer Glasses}}}\ (\bibinfo
	{publisher} {CRC Press (Boca Raton, FL, USA)},\ \bibinfo {year}
	{2017})\BibitemShut {NoStop}%
	\bibitem [{\citenamefont {Ruta}\ \emph {et~al.}(2017)\citenamefont {Ruta},
		\citenamefont {Pineda},\ and\ \citenamefont {Evenson}}]{rut17}%
	\BibitemOpen
	\bibfield  {author} {\bibinfo {author} {\bibfnamefont {B.}~\bibnamefont
			{Ruta}}, \bibinfo {author} {\bibfnamefont {E.}~\bibnamefont {Pineda}},\ and\
		\bibinfo {author} {\bibfnamefont {Z.}~\bibnamefont {Evenson}},\ }\bibfield
	{title} {\bibinfo {title} {Relaxation processes and physical aging in
			metallic glasses},\ }\href {https://doi.org/10.1088/1361-648X/aa9964}
	{\bibfield  {journal} {\bibinfo  {journal} {J. Phys.: Condens. Matter}\
		}\textbf {\bibinfo {volume} {29}},\ \bibinfo {pages} {503002} (\bibinfo
		{year} {2017})}\BibitemShut {NoStop}%
	\bibitem [{\citenamefont {Mauro}(2021)}]{mauro}%
	\BibitemOpen
	\bibfield  {author} {\bibinfo {author} {\bibfnamefont {J.~C.}\ \bibnamefont
			{Mauro}},\ }\href@noop {} {\emph {\bibinfo {title} {{Materials Kinetics:
					Transport and Rate Phenomena}}}}\ (\bibinfo  {publisher} {Elsevier,
		Amsterdan, Netherlands},\ \bibinfo {year} {2021})\BibitemShut {NoStop}%
	\bibitem [{\citenamefont {Tool}(1946)}]{too46}%
	\BibitemOpen
	\bibfield  {author} {\bibinfo {author} {\bibfnamefont {A.~Q.}\ \bibnamefont
			{Tool}},\ }\bibfield  {title} {\bibinfo {title} {Relation between inelastic
			deformability and thermal expansion of glass in its annealing range},\
	}\href@noop {} {\bibfield  {journal} {\bibinfo  {journal} {J. Amer. Ceram.
				Soc.}\ }\textbf {\bibinfo {volume} {29}},\ \bibinfo {pages} {240} (\bibinfo
		{year} {1946})}\BibitemShut {NoStop}%
	\bibitem [{\citenamefont {Kovacs}(1963)}]{kov63}%
	\BibitemOpen
	\bibfield  {author} {\bibinfo {author} {\bibfnamefont {A.~J.}\ \bibnamefont
			{Kovacs}},\ }\bibfield  {title} {\bibinfo {title} {Transition vitreuse dans
			les polymeres amorphes. {Etude} phenomenologique},\ }\href@noop {} {\bibfield
		{journal} {\bibinfo  {journal} {Fortschr. Hochpolym.-Forsch.}\ }\textbf
		{\bibinfo {volume} {3}},\ \bibinfo {pages} {394} (\bibinfo {year}
		{1963})}\BibitemShut {NoStop}%
	\bibitem [{\citenamefont {Moynihan}\ \emph {et~al.}(1976)\citenamefont
		{Moynihan}, \citenamefont {Macedo}, \citenamefont {Montrose}, \citenamefont
		{Gupta}, \citenamefont {DeBolt}, \citenamefont {Dill}, \citenamefont {Dom},
		\citenamefont {Drake}, \citenamefont {Easteal}, \citenamefont {Elterman},
		\citenamefont {Moeller}, \citenamefont {Sasabe},\ and\ \citenamefont
		{Wilder}}]{moy76a}%
	\BibitemOpen
	\bibfield  {author} {\bibinfo {author} {\bibfnamefont {C.~T.}\ \bibnamefont
			{Moynihan}}, \bibinfo {author} {\bibfnamefont {P.~B.}\ \bibnamefont
			{Macedo}}, \bibinfo {author} {\bibfnamefont {C.~J.}\ \bibnamefont
			{Montrose}}, \bibinfo {author} {\bibfnamefont {P.~K.}\ \bibnamefont {Gupta}},
		\bibinfo {author} {\bibfnamefont {M.~A.}\ \bibnamefont {DeBolt}}, \bibinfo
		{author} {\bibfnamefont {J.~F.}\ \bibnamefont {Dill}}, \bibinfo {author}
		{\bibfnamefont {B.~E.}\ \bibnamefont {Dom}}, \bibinfo {author} {\bibfnamefont
			{P.~W.}\ \bibnamefont {Drake}}, \bibinfo {author} {\bibfnamefont {A.~J.}\
			\bibnamefont {Easteal}}, \bibinfo {author} {\bibfnamefont {P.~B.}\
			\bibnamefont {Elterman}}, \bibinfo {author} {\bibfnamefont {R.~P.}\
			\bibnamefont {Moeller}}, \bibinfo {author} {\bibfnamefont {H.}~\bibnamefont
			{Sasabe}},\ and\ \bibinfo {author} {\bibfnamefont {J.~A.}\ \bibnamefont
			{Wilder}},\ }\bibfield  {title} {\bibinfo {title} {Structural relaxation in
			vitreous materials},\ }\href@noop {} {\bibfield  {journal} {\bibinfo
			{journal} {Ann. NY Acad. Sci.}\ }\textbf {\bibinfo {volume} {279}},\ \bibinfo
		{pages} {15} (\bibinfo {year} {1976})}\BibitemShut {NoStop}%
	\bibitem [{\citenamefont {Chen}(1978)}]{che78}%
	\BibitemOpen
	\bibfield  {author} {\bibinfo {author} {\bibfnamefont {H.~S.}\ \bibnamefont
			{Chen}},\ }\bibfield  {title} {\bibinfo {title} {The influence of structural
			relaxation on the density and {Young's} modulus of metallic glasses},\ }\href
	{https://doi.org/http://dx.doi.org/10.1063/1.325279} {\bibfield  {journal}
		{\bibinfo  {journal} {J. Appl. Phys.}\ }\textbf {\bibinfo {volume} {49}},\
		\bibinfo {pages} {3289} (\bibinfo {year} {1978})}\BibitemShut {NoStop}%
	\bibitem [{\citenamefont {Olsen}\ \emph {et~al.}(1998)\citenamefont {Olsen},
		\citenamefont {Dyre},\ and\ \citenamefont {Christensen}}]{ols98}%
	\BibitemOpen
	\bibfield  {author} {\bibinfo {author} {\bibfnamefont {N.~B.}\ \bibnamefont
			{Olsen}}, \bibinfo {author} {\bibfnamefont {J.~C.}\ \bibnamefont {Dyre}},\
		and\ \bibinfo {author} {\bibfnamefont {T.}~\bibnamefont {Christensen}},\
	}\bibfield  {title} {\bibinfo {title} {Structural relaxation monitored by
			instantaneous shear modulus},\ }\href
	{https://doi.org/10.1103/PhysRevLett.81.1031} {\bibfield  {journal} {\bibinfo
			{journal} {Phys. Rev. Lett.}\ }\textbf {\bibinfo {volume} {81}},\ \bibinfo
		{pages} {1031} (\bibinfo {year} {1998})}\BibitemShut {NoStop}%
	\bibitem [{\citenamefont {Dyre}\ and\ \citenamefont {Olsen}(2003)}]{dyr03}%
	\BibitemOpen
	\bibfield  {author} {\bibinfo {author} {\bibfnamefont {J.~C.}\ \bibnamefont
			{Dyre}}\ and\ \bibinfo {author} {\bibfnamefont {N.~B.}\ \bibnamefont
			{Olsen}},\ }\bibfield  {title} {\bibinfo {title} {Minimal model for beta
			relaxation in viscous liquids},\ }\href
	{https://doi.org/10.1103/PhysRevLett.91.155703} {\bibfield  {journal}
		{\bibinfo  {journal} {Phys. Rev. Lett.}\ }\textbf {\bibinfo {volume} {91}},\
		\bibinfo {pages} {155703} (\bibinfo {year} {2003})}\BibitemShut {NoStop}%
	\bibitem [{\citenamefont {Di~Leonardo}\ \emph {et~al.}(2004)\citenamefont
		{Di~Leonardo}, \citenamefont {Scopigno}, \citenamefont {Ruocco},\ and\
		\citenamefont {Buontempo}}]{dil04}%
	\BibitemOpen
	\bibfield  {author} {\bibinfo {author} {\bibfnamefont {R.}~\bibnamefont
			{Di~Leonardo}}, \bibinfo {author} {\bibfnamefont {T.}~\bibnamefont
			{Scopigno}}, \bibinfo {author} {\bibfnamefont {G.}~\bibnamefont {Ruocco}},\
		and\ \bibinfo {author} {\bibfnamefont {U.}~\bibnamefont {Buontempo}},\
	}\bibfield  {title} {\bibinfo {title} {Spectroscopic cell for fast pressure
			jumps across the glass transition line},\ }\href
	{https://doi.org/http://dx.doi.org/10.1063/1.1763253} {\bibfield  {journal}
		{\bibinfo  {journal} {Rev. Sci. Instrum.}\ }\textbf {\bibinfo {volume}
			{75}},\ \bibinfo {pages} {2631} (\bibinfo {year} {2004})}\BibitemShut
	{NoStop}%
	\bibitem [{\citenamefont {Schlosser}\ and\ \citenamefont
		{Sch{\"o}nhals}(1991)}]{sch91}%
	\BibitemOpen
	\bibfield  {author} {\bibinfo {author} {\bibfnamefont {E.}~\bibnamefont
			{Schlosser}}\ and\ \bibinfo {author} {\bibfnamefont {A.}~\bibnamefont
			{Sch{\"o}nhals}},\ }\bibfield  {title} {\bibinfo {title} {Dielectric
			relaxation during physical aging},\ }\href@noop {} {\bibfield  {journal}
		{\bibinfo  {journal} {Polymer}\ }\textbf {\bibinfo {volume} {32}},\ \bibinfo
		{pages} {2135} (\bibinfo {year} {1991})}\BibitemShut {NoStop}%
	\bibitem [{\citenamefont {Lunkenheimer}\ \emph {et~al.}(2005)\citenamefont
		{Lunkenheimer}, \citenamefont {Wehn}, \citenamefont {Schneider},\ and\
		\citenamefont {Loidl}}]{lun05}%
	\BibitemOpen
	\bibfield  {author} {\bibinfo {author} {\bibfnamefont {P.}~\bibnamefont
			{Lunkenheimer}}, \bibinfo {author} {\bibfnamefont {R.}~\bibnamefont {Wehn}},
		\bibinfo {author} {\bibfnamefont {U.}~\bibnamefont {Schneider}},\ and\
		\bibinfo {author} {\bibfnamefont {A.}~\bibnamefont {Loidl}},\ }\bibfield
	{title} {\bibinfo {title} {Glassy aging dynamics},\ }\href@noop {} {\bibfield
		{journal} {\bibinfo  {journal} {Phys. Rev. Lett.}\ }\textbf {\bibinfo
			{volume} {95}},\ \bibinfo {pages} {055702} (\bibinfo {year}
		{2005})}\BibitemShut {NoStop}%
	\bibitem [{\citenamefont {Brun}\ \emph {et~al.}(2012)\citenamefont {Brun},
		\citenamefont {Ladieu}, \citenamefont {L'Hote}, \citenamefont {Biroli},\ and\
		\citenamefont {Bouchaud}}]{bru12}%
	\BibitemOpen
	\bibfield  {author} {\bibinfo {author} {\bibfnamefont {C.}~\bibnamefont
			{Brun}}, \bibinfo {author} {\bibfnamefont {F.}~\bibnamefont {Ladieu}},
		\bibinfo {author} {\bibfnamefont {D.}~\bibnamefont {L'Hote}}, \bibinfo
		{author} {\bibfnamefont {G.}~\bibnamefont {Biroli}},\ and\ \bibinfo {author}
		{\bibfnamefont {J.-P.}\ \bibnamefont {Bouchaud}},\ }\bibfield  {title}
	{\bibinfo {title} {Evidence of growing spatial correlations during the aging
			of glassy glycerol},\ }\href {https://doi.org/10.1103/PhysRevLett.109.175702}
	{\bibfield  {journal} {\bibinfo  {journal} {Phys. Rev. Lett.}\ }\textbf
		{\bibinfo {volume} {109}},\ \bibinfo {pages} {175702} (\bibinfo {year}
		{2012})}\BibitemShut {NoStop}%
	\bibitem [{\citenamefont {Richert}(2015)}]{ric15}%
	\BibitemOpen
	\bibfield  {author} {\bibinfo {author} {\bibfnamefont {R.}~\bibnamefont
			{Richert}},\ }\bibfield  {title} {\bibinfo {title} {Supercooled liquids and
			glasses by dielectric relaxation spectroscopy},\ }\href
	{https://doi.org/10.1002/9781118949702.ch4} {\bibfield  {journal} {\bibinfo
			{journal} {Adv. Chem. Phys.}\ }\textbf {\bibinfo {volume} {156}},\ \bibinfo
		{pages} {101} (\bibinfo {year} {2015})}\BibitemShut {NoStop}%
	\bibitem [{\citenamefont {{Di Lisio}}\ \emph {et~al.}(2024)\citenamefont {{Di
				Lisio}}, \citenamefont {Rocchi},\ and\ \citenamefont {Cangialosi}}]{dil24}%
	\BibitemOpen
	\bibfield  {author} {\bibinfo {author} {\bibfnamefont {V.}~\bibnamefont {{Di
					Lisio}}}, \bibinfo {author} {\bibfnamefont {L.~A.}\ \bibnamefont {Rocchi}},\
		and\ \bibinfo {author} {\bibfnamefont {D.}~\bibnamefont {Cangialosi}},\
	}\bibfield  {title} {\bibinfo {title} {Twofold facet of kinetics of glass
			aging},\ }\href {https://doi.org/10.1103/PhysRevLett.133.048201} {\bibfield
		{journal} {\bibinfo  {journal} {Phys. Rev. Lett.}\ }\textbf {\bibinfo
			{volume} {133}},\ \bibinfo {pages} {048201} (\bibinfo {year}
		{2024})}\BibitemShut {NoStop}%
	\bibitem [{\citenamefont {Bera}\ \emph {et~al.}(2024)\citenamefont {Bera},
		\citenamefont {Medvedev}, \citenamefont {Caruthers},\ and\ \citenamefont
		{Ediger}}]{ber24}%
	\BibitemOpen
	\bibfield  {author} {\bibinfo {author} {\bibfnamefont {P.~K.}\ \bibnamefont
			{Bera}}, \bibinfo {author} {\bibfnamefont {G.~A.}\ \bibnamefont {Medvedev}},
		\bibinfo {author} {\bibfnamefont {J.~M.}\ \bibnamefont {Caruthers}},\ and\
		\bibinfo {author} {\bibfnamefont {M.~D.}\ \bibnamefont {Ediger}},\ }\bibfield
	{title} {\bibinfo {title} {Structural relaxation time of a polymer glass
			during deformation},\ }\href {https://doi.org/10.1103/PhysRevLett.132.208101}
	{\bibfield  {journal} {\bibinfo  {journal} {Phys. Rev. Lett.}\ }\textbf
		{\bibinfo {volume} {132}},\ \bibinfo {pages} {208101} (\bibinfo {year}
		{2024})}\BibitemShut {NoStop}%
	\bibitem [{\citenamefont {Igarashi}\ \emph {et~al.}(2008)\citenamefont
		{Igarashi}, \citenamefont {Christensen}, \citenamefont {Larsen},
		\citenamefont {Olsen}, \citenamefont {Pedersen}, \citenamefont {Rasmussen},\
		and\ \citenamefont {Dyre}}]{iga08a}%
	\BibitemOpen
	\bibfield  {author} {\bibinfo {author} {\bibfnamefont {B.}~\bibnamefont
			{Igarashi}}, \bibinfo {author} {\bibfnamefont {T.}~\bibnamefont
			{Christensen}}, \bibinfo {author} {\bibfnamefont {E.~H.}\ \bibnamefont
			{Larsen}}, \bibinfo {author} {\bibfnamefont {N.~B.}\ \bibnamefont {Olsen}},
		\bibinfo {author} {\bibfnamefont {I.~H.}\ \bibnamefont {Pedersen}}, \bibinfo
		{author} {\bibfnamefont {T.}~\bibnamefont {Rasmussen}},\ and\ \bibinfo
		{author} {\bibfnamefont {J.~C.}\ \bibnamefont {Dyre}},\ }\bibfield  {title}
	{\bibinfo {title} {A cryostat and temperature control system optimized for
			measuring relaxations of glass-forming liquids},\ }\href
	{https://doi.org/http://dx.doi.org/10.1063/1.2903419} {\bibfield  {journal}
		{\bibinfo  {journal} {Rev. Sci. Instrum.}\ }\textbf {\bibinfo {volume}
			{79}},\ \bibinfo {pages} {045105} (\bibinfo {year} {2008})}\BibitemShut
	{NoStop}%
	\bibitem [{\citenamefont {Hénot}\ and\ \citenamefont
		{Ladieu}(2023)}]{henot2023}%
	\BibitemOpen
	\bibfield  {author} {\bibinfo {author} {\bibfnamefont {M.}~\bibnamefont
			{Hénot}}\ and\ \bibinfo {author} {\bibfnamefont {F.}~\bibnamefont
			{Ladieu}},\ }\bibfield  {title} {\bibinfo {title} {Non-linear physical aging
			of supercooled glycerol induced by large upward ideal temperature steps
			monitored through cooling experiments},\ }\href@noop {} {\bibfield  {journal}
		{\bibinfo  {journal} {J. Chem. Phys.}\ }\textbf {\bibinfo {volume} {158}},\
		\bibinfo {pages} {224504} (\bibinfo {year} {2023})}\BibitemShut {NoStop}%
	\bibitem [{\citenamefont {Di~Lisio}\ \emph {et~al.}(2023)\citenamefont
		{Di~Lisio}, \citenamefont {Stavropoulou},\ and\ \citenamefont
		{Cangialosi}}]{dilisio2023}%
	\BibitemOpen
	\bibfield  {author} {\bibinfo {author} {\bibfnamefont {V.}~\bibnamefont
			{Di~Lisio}}, \bibinfo {author} {\bibfnamefont {V.-M.}\ \bibnamefont
			{Stavropoulou}},\ and\ \bibinfo {author} {\bibfnamefont {D.}~\bibnamefont
			{Cangialosi}},\ }\bibfield  {title} {\bibinfo {title} {Physical aging in
			molecular glasses beyond the $\alpha$ relaxation},\ }\href@noop {} {\bibfield
		{journal} {\bibinfo  {journal} {J. Chem. Phys.}\ }\textbf {\bibinfo {volume}
			{159}},\ \bibinfo {pages} {064505} (\bibinfo {year} {2023})}\BibitemShut
	{NoStop}%
	\bibitem [{\citenamefont {Di~Lisio}\ \emph {et~al.}(2024)\citenamefont
		{Di~Lisio}, \citenamefont {Rocchi},\ and\ \citenamefont
		{Cangialosi}}]{dilisio2024}%
	\BibitemOpen
	\bibfield  {author} {\bibinfo {author} {\bibfnamefont {V.}~\bibnamefont
			{Di~Lisio}}, \bibinfo {author} {\bibfnamefont {L.~A.}\ \bibnamefont
			{Rocchi}},\ and\ \bibinfo {author} {\bibfnamefont {D.}~\bibnamefont
			{Cangialosi}},\ }\bibfield  {title} {\bibinfo {title} {Twofold facet of
			kinetics of glass aging},\ }\href
	{https://doi.org/10.1103/PhysRevLett.133.048201} {\bibfield  {journal}
		{\bibinfo  {journal} {Phys. Rev. Lett.}\ }\textbf {\bibinfo {volume} {133}},\
		\bibinfo {pages} {048201} (\bibinfo {year} {2024})}\BibitemShut {NoStop}%
	\bibitem [{\citenamefont {Mauro}\ \emph {et~al.}(2009)\citenamefont {Mauro},
		\citenamefont {Gupta},\ and\ \citenamefont {Loucks}}]{mau09}%
	\BibitemOpen
	\bibfield  {author} {\bibinfo {author} {\bibfnamefont {J.~C.}\ \bibnamefont
			{Mauro}}, \bibinfo {author} {\bibfnamefont {P.~K.}\ \bibnamefont {Gupta}},\
		and\ \bibinfo {author} {\bibfnamefont {R.~J.}\ \bibnamefont {Loucks}},\
	}\bibfield  {title} {\bibinfo {title} {Composition dependence of glass
			transition temperature and fragility. {II. A} topological model of alkali
			borate liquids},\ }\href@noop {} {\bibfield  {journal} {\bibinfo  {journal}
			{J. Chem. Phys.}\ }\textbf {\bibinfo {volume} {130}},\ \bibinfo {pages}
		{234503} (\bibinfo {year} {2009})}\BibitemShut {NoStop}%
	\bibitem [{\citenamefont {Nielsen}\ and\ \citenamefont {Dyre}(1996)}]{nie96}%
	\BibitemOpen
	\bibfield  {author} {\bibinfo {author} {\bibfnamefont {J.~K.}\ \bibnamefont
			{Nielsen}}\ and\ \bibinfo {author} {\bibfnamefont {J.~C.}\ \bibnamefont
			{Dyre}},\ }\bibfield  {title} {\bibinfo {title} {Fluctuation-dissipation
			theorem for frequency-dependent specific heat},\ }\href
	{https://doi.org/10.1103/PhysRevB.54.15754} {\bibfield  {journal} {\bibinfo
			{journal} {Phys. Rev. B}\ }\textbf {\bibinfo {volume} {54}},\ \bibinfo
		{pages} {15754} (\bibinfo {year} {1996})}\BibitemShut {NoStop}%
	\bibitem [{\citenamefont {Niss}\ \emph {et~al.}(2020)\citenamefont {Niss},
		\citenamefont {Dyre},\ and\ \citenamefont {Hecksher}}]{nis20}%
	\BibitemOpen
	\bibfield  {author} {\bibinfo {author} {\bibfnamefont {K.}~\bibnamefont
			{Niss}}, \bibinfo {author} {\bibfnamefont {J.~C.}\ \bibnamefont {Dyre}},\
		and\ \bibinfo {author} {\bibfnamefont {T.}~\bibnamefont {Hecksher}},\
	}\bibfield  {title} {\bibinfo {title} {Long-time structural relaxation of
			glass-forming liquids: {Simple} or stretched exponential?},\ }\href
	{https://doi.org/10.1063/1.5142189} {\bibfield  {journal} {\bibinfo
			{journal} {J. Chem. Phys.}\ }\textbf {\bibinfo {volume} {152}},\ \bibinfo
		{pages} {041103} (\bibinfo {year} {2020})}\BibitemShut {NoStop}%
	\bibitem [{\citenamefont {Riechers}\ \emph {et~al.}(2022)\citenamefont
		{Riechers}, \citenamefont {Roed}, \citenamefont {Mehri}, \citenamefont
		{Ingebrigtsen}, \citenamefont {Hecksher}, \citenamefont {Dyre},\ and\
		\citenamefont {Niss}}]{rie22}%
	\BibitemOpen
	\bibfield  {author} {\bibinfo {author} {\bibfnamefont {B.}~\bibnamefont
			{Riechers}}, \bibinfo {author} {\bibfnamefont {L.~A.}\ \bibnamefont {Roed}},
		\bibinfo {author} {\bibfnamefont {S.}~\bibnamefont {Mehri}}, \bibinfo
		{author} {\bibfnamefont {T.~S.}\ \bibnamefont {Ingebrigtsen}}, \bibinfo
		{author} {\bibfnamefont {T.}~\bibnamefont {Hecksher}}, \bibinfo {author}
		{\bibfnamefont {J.~C.}\ \bibnamefont {Dyre}},\ and\ \bibinfo {author}
		{\bibfnamefont {K.}~\bibnamefont {Niss}},\ }\bibfield  {title} {\bibinfo
		{title} {Predicting nonlinear physical aging of glasses from equilibrium
			relaxation via the material time},\ }\href
	{https://doi.org/10.1126/sciadv.abl9809} {\bibfield  {journal} {\bibinfo
			{journal} {Sci. Adv.}\ }\textbf {\bibinfo {volume} {8}},\ \bibinfo {pages}
		{eabl9809} (\bibinfo {year} {2022})}\BibitemShut {NoStop}%
	\bibitem [{\citenamefont {Hecksher}\ \emph {et~al.}(2015)\citenamefont
		{Hecksher}, \citenamefont {Olsen},\ and\ \citenamefont {Dyre}}]{hec15}%
	\BibitemOpen
	\bibfield  {author} {\bibinfo {author} {\bibfnamefont {T.}~\bibnamefont
			{Hecksher}}, \bibinfo {author} {\bibfnamefont {N.~B.}\ \bibnamefont
			{Olsen}},\ and\ \bibinfo {author} {\bibfnamefont {J.~C.}\ \bibnamefont
			{Dyre}},\ }\bibfield  {title} {\bibinfo {title} {Communication: Direct tests
			of single-parameter aging},\ }\href
	{https://doi.org/http://dx.doi.org/10.1063/1.4923000} {\bibfield  {journal}
		{\bibinfo  {journal} {J. Chem. Phys.}\ }\textbf {\bibinfo {volume} {142}},\
		\bibinfo {pages} {241103} (\bibinfo {year} {2015})}\BibitemShut {NoStop}%
	\bibitem [{\citenamefont {B{\"o}hmer}\ \emph {et~al.}(2014)\citenamefont
		{B{\"o}hmer}, \citenamefont {Gainaru},\ and\ \citenamefont
		{Richert}}]{boh14a}%
	\BibitemOpen
	\bibfield  {author} {\bibinfo {author} {\bibfnamefont {R.}~\bibnamefont
			{B{\"o}hmer}}, \bibinfo {author} {\bibfnamefont {C.}~\bibnamefont
			{Gainaru}},\ and\ \bibinfo {author} {\bibfnamefont {R.}~\bibnamefont
			{Richert}},\ }\bibfield  {title} {\bibinfo {title} {Structure and dynamics of
			monohydroxy alcohols -- {Milestones} towards their microscopic
			understanding{, 100}\, years after {Debye}},\ }\href
	{https://doi.org/http://dx.doi.org/10.1016/j.physrep.2014.07.005} {\bibfield
		{journal} {\bibinfo  {journal} {Phys. Rep.}\ }\textbf {\bibinfo {volume}
			{545}},\ \bibinfo {pages} {125 } (\bibinfo {year} {2014})}\BibitemShut
	{NoStop}%
	\bibitem [{\citenamefont {Huth}\ \emph {et~al.}(2007)\citenamefont {Huth},
		\citenamefont {Wang}, \citenamefont {Schick},\ and\ \citenamefont
		{Richert}}]{huth2007comparing}%
	\BibitemOpen
	\bibfield  {author} {\bibinfo {author} {\bibfnamefont {H.}~\bibnamefont
			{Huth}}, \bibinfo {author} {\bibfnamefont {L.-M.}\ \bibnamefont {Wang}},
		\bibinfo {author} {\bibfnamefont {C.}~\bibnamefont {Schick}},\ and\ \bibinfo
		{author} {\bibfnamefont {R.}~\bibnamefont {Richert}},\ }\bibfield  {title}
	{\bibinfo {title} {Comparing calorimetric and dielectric polarization modes
			in viscous 2-ethyl-1-hexanol},\ }\href@noop {} {\bibfield  {journal}
		{\bibinfo  {journal} {J. Chem. Phys.}\ }\textbf {\bibinfo {volume} {126}},\
		\bibinfo {pages} {104503} (\bibinfo {year} {2007})}\BibitemShut {NoStop}%
	\bibitem [{\citenamefont {Wang}\ \emph {et~al.}(2008)\citenamefont {Wang},
		\citenamefont {Tian}, \citenamefont {Liu},\ and\ \citenamefont
		{Richert}}]{wang2008calorimetric}%
	\BibitemOpen
	\bibfield  {author} {\bibinfo {author} {\bibfnamefont {L.-M.}\ \bibnamefont
			{Wang}}, \bibinfo {author} {\bibfnamefont {Y.}~\bibnamefont {Tian}}, \bibinfo
		{author} {\bibfnamefont {R.}~\bibnamefont {Liu}},\ and\ \bibinfo {author}
		{\bibfnamefont {R.}~\bibnamefont {Richert}},\ }\bibfield  {title} {\bibinfo
		{title} {Calorimetric versus kinetic glass transitions in viscous monohydroxy
			alcohols},\ }\href@noop {} {\bibfield  {journal} {\bibinfo  {journal} {J.
				Chem. Phys.}\ }\textbf {\bibinfo {volume} {128}},\ \bibinfo {pages} {084503}
		(\bibinfo {year} {2008})}\BibitemShut {NoStop}%
	\bibitem [{\citenamefont {Cutroni}\ and\ \citenamefont
		{Mandanici}(2001)}]{cutroni2001}%
	\BibitemOpen
	\bibfield  {author} {\bibinfo {author} {\bibfnamefont {M.}~\bibnamefont
			{Cutroni}}\ and\ \bibinfo {author} {\bibfnamefont {A.}~\bibnamefont
			{Mandanici}},\ }\bibfield  {title} {\bibinfo {title} {The $\alpha$-relaxation
			process in simple glass forming liquid m-toluidine. {II. The} temperature
			dependence of the mechanical response},\ }\href
	{https://doi.org/10.1063/1.1358878} {\bibfield  {journal} {\bibinfo
			{journal} {J. Chem. Phys.}\ }\textbf {\bibinfo {volume} {114}},\ \bibinfo
		{pages} {7124} (\bibinfo {year} {2001})}\BibitemShut {NoStop}%
	\bibitem [{\citenamefont {Jakobsen}\ \emph {et~al.}(2005)\citenamefont
		{Jakobsen}, \citenamefont {Niss},\ and\ \citenamefont
		{Olsen}}]{jakobsen2005}%
	\BibitemOpen
	\bibfield  {author} {\bibinfo {author} {\bibfnamefont {B.}~\bibnamefont
			{Jakobsen}}, \bibinfo {author} {\bibfnamefont {K.}~\bibnamefont {Niss}},\
		and\ \bibinfo {author} {\bibfnamefont {N.~B.}\ \bibnamefont {Olsen}},\
	}\bibfield  {title} {\bibinfo {title} {Dielectric and shear mechanical alpha
			and beta relaxations in seven glass-forming liquids},\ }\href
	{https://doi.org/10.1063/1.2136887} {\bibfield  {journal} {\bibinfo
			{journal} {J. Chem. Phys.}\ }\textbf {\bibinfo {volume} {123}},\ \bibinfo
		{pages} {234511} (\bibinfo {year} {2005})}\BibitemShut {NoStop}%
	\bibitem [{\citenamefont {Jakobsen}\ \emph {et~al.}(2012)\citenamefont
		{Jakobsen}, \citenamefont {Hecksher}, \citenamefont {Christensen},
		\citenamefont {Olsen}, \citenamefont {Dyre},\ and\ \citenamefont
		{Niss}}]{jakobsen2012communication}%
	\BibitemOpen
	\bibfield  {author} {\bibinfo {author} {\bibfnamefont {B.}~\bibnamefont
			{Jakobsen}}, \bibinfo {author} {\bibfnamefont {T.}~\bibnamefont {Hecksher}},
		\bibinfo {author} {\bibfnamefont {T.}~\bibnamefont {Christensen}}, \bibinfo
		{author} {\bibfnamefont {N.~B.}\ \bibnamefont {Olsen}}, \bibinfo {author}
		{\bibfnamefont {J.~C.}\ \bibnamefont {Dyre}},\ and\ \bibinfo {author}
		{\bibfnamefont {K.}~\bibnamefont {Niss}},\ }\bibfield  {title} {\bibinfo
		{title} {Communication: Identical temperature dependence of the time scales
			of several linear-response functions of two glass-forming liquids},\
	}\href@noop {} {\bibfield  {journal} {\bibinfo  {journal} {J. Chem. Phys.}\
		}\textbf {\bibinfo {volume} {136}},\ \bibinfo {pages} {081102} (\bibinfo
		{year} {2012})}\BibitemShut {NoStop}%
	\bibitem [{\citenamefont {Roed}\ \emph {et~al.}(2021)\citenamefont {Roed},
		\citenamefont {Dyre}, \citenamefont {Niss}, \citenamefont {Hecksher},\ and\
		\citenamefont {Riechers}}]{roe21}%
	\BibitemOpen
	\bibfield  {author} {\bibinfo {author} {\bibfnamefont {L.~A.}\ \bibnamefont
			{Roed}}, \bibinfo {author} {\bibfnamefont {J.~C.}\ \bibnamefont {Dyre}},
		\bibinfo {author} {\bibfnamefont {K.}~\bibnamefont {Niss}}, \bibinfo {author}
		{\bibfnamefont {T.}~\bibnamefont {Hecksher}},\ and\ \bibinfo {author}
		{\bibfnamefont {B.}~\bibnamefont {Riechers}},\ }\bibfield  {title} {\bibinfo
		{title} {Time-scale ordering in hydrogen- and van der {Waals-bonded}
			liquids},\ }\href {https://doi.org/10.1063/5.0049108} {\bibfield  {journal}
		{\bibinfo  {journal} {J. Chem. Phys.}\ }\textbf {\bibinfo {volume} {154}},\
		\bibinfo {pages} {184508} (\bibinfo {year} {2021})}\BibitemShut {NoStop}%
	\bibitem [{\citenamefont {Gainaru}\ and\ \citenamefont
		{B{\"o}hmer}(2010)}]{gainaru2010coupling}%
	\BibitemOpen
	\bibfield  {author} {\bibinfo {author} {\bibfnamefont {C.}~\bibnamefont
			{Gainaru}}\ and\ \bibinfo {author} {\bibfnamefont {R.}~\bibnamefont
			{B{\"o}hmer}},\ }\bibfield  {title} {\bibinfo {title} {Coupling of the
			electrical conductivity to the structural relaxation, absence of physical
			aging on the time scale of the {Debye} process, and number of correlated
			molecules in the supercooled monohydroxy alcohol 2-ethylhexanol},\
	}\href@noop {} {\bibfield  {journal} {\bibinfo  {journal} {J. Non-Cryst.
				Solids}\ }\textbf {\bibinfo {volume} {356}},\ \bibinfo {pages} {542}
		(\bibinfo {year} {2010})}\BibitemShut {NoStop}%
	\bibitem [{\citenamefont {Moch}\ \emph {et~al.}(2022)\citenamefont {Moch},
		\citenamefont {M{\"u}nzner}, \citenamefont {B{\"o}hmer},\ and\ \citenamefont
		{Gainaru}}]{moch2022molecular}%
	\BibitemOpen
	\bibfield  {author} {\bibinfo {author} {\bibfnamefont {K.}~\bibnamefont
			{Moch}}, \bibinfo {author} {\bibfnamefont {P.}~\bibnamefont {M{\"u}nzner}},
		\bibinfo {author} {\bibfnamefont {R.}~\bibnamefont {B{\"o}hmer}},\ and\
		\bibinfo {author} {\bibfnamefont {C.}~\bibnamefont {Gainaru}},\ }\bibfield
	{title} {\bibinfo {title} {Molecular cross-correlations govern structural
			rearrangements in a nonassociating polar glass former},\ }\href@noop {}
	{\bibfield  {journal} {\bibinfo  {journal} {Physical Review Letters}\
		}\textbf {\bibinfo {volume} {128}},\ \bibinfo {pages} {228001} (\bibinfo
		{year} {2022})}\BibitemShut {NoStop}%
	\bibitem [{\citenamefont {Gabriel}\ and\ \citenamefont
		{Richert}(2023)}]{gabriel2023comparing}%
	\BibitemOpen
	\bibfield  {author} {\bibinfo {author} {\bibfnamefont {J.~P.}\ \bibnamefont
			{Gabriel}}\ and\ \bibinfo {author} {\bibfnamefont {R.}~\bibnamefont
			{Richert}},\ }\bibfield  {title} {\bibinfo {title} {Comparing two sources of
			physical aging: Temperature vs electric field},\ }\href@noop {} {\bibfield
		{journal} {\bibinfo  {journal} {J. Chem. Phys.}\ }\textbf {\bibinfo {volume}
			{159}},\ \bibinfo {pages} {164502} (\bibinfo {year} {2023})}\BibitemShut
	{NoStop}%
	\bibitem [{\citenamefont {Preus}\ \emph {et~al.}(2012)\citenamefont {Preus},
		\citenamefont {Gainaru}, \citenamefont {Hecksher}, \citenamefont {Bauer},
		\citenamefont {Dyre}, \citenamefont {Richert},\ and\ \citenamefont
		{B{\"o}hmer}}]{pre12}%
	\BibitemOpen
	\bibfield  {author} {\bibinfo {author} {\bibfnamefont {M.}~\bibnamefont
			{Preus}}, \bibinfo {author} {\bibfnamefont {C.}~\bibnamefont {Gainaru}},
		\bibinfo {author} {\bibfnamefont {T.}~\bibnamefont {Hecksher}}, \bibinfo
		{author} {\bibfnamefont {S.}~\bibnamefont {Bauer}}, \bibinfo {author}
		{\bibfnamefont {J.~C.}\ \bibnamefont {Dyre}}, \bibinfo {author}
		{\bibfnamefont {R.}~\bibnamefont {Richert}},\ and\ \bibinfo {author}
		{\bibfnamefont {R.}~\bibnamefont {B{\"o}hmer}},\ }\bibfield  {title}
	{\bibinfo {title} {Experimental studies of {Debye-like} process and
			structural relaxation in mixtures of 2-ethyl-1-hexanol and 2-ethyl-1-hexyl
			bromide},\ }\href {https://doi.org/10.1063/1.4755754} {\bibfield  {journal}
		{\bibinfo  {journal} {J. Chem. Phys.}\ }\textbf {\bibinfo {volume} {137}},\
		\bibinfo {pages} {144502} (\bibinfo {year} {2012})}\BibitemShut {NoStop}%
	\bibitem [{\citenamefont {Gainaru}\ \emph {et~al.}(2011)\citenamefont
		{Gainaru}, \citenamefont {Kastner}, \citenamefont {Mayr}, \citenamefont
		{Lunkenheimer}, \citenamefont {Schildmann}, \citenamefont {Weber},
		\citenamefont {Hiller}, \citenamefont {Loidl},\ and\ \citenamefont
		{B{\"o}hmer}}]{gai11}%
	\BibitemOpen
	\bibfield  {author} {\bibinfo {author} {\bibfnamefont {C.}~\bibnamefont
			{Gainaru}}, \bibinfo {author} {\bibfnamefont {S.}~\bibnamefont {Kastner}},
		\bibinfo {author} {\bibfnamefont {F.}~\bibnamefont {Mayr}}, \bibinfo {author}
		{\bibfnamefont {P.}~\bibnamefont {Lunkenheimer}}, \bibinfo {author}
		{\bibfnamefont {S.}~\bibnamefont {Schildmann}}, \bibinfo {author}
		{\bibfnamefont {H.~J.}\ \bibnamefont {Weber}}, \bibinfo {author}
		{\bibfnamefont {W.}~\bibnamefont {Hiller}}, \bibinfo {author} {\bibfnamefont
			{A.}~\bibnamefont {Loidl}},\ and\ \bibinfo {author} {\bibfnamefont
			{R.}~\bibnamefont {B{\"o}hmer}},\ }\bibfield  {title} {\bibinfo {title}
		{Hydrogen-bond equilibria and lifetimes in a monohydroxy alcohol},\ }\href
	{https://doi.org/10.1103/PhysRevLett.107.118304} {\bibfield  {journal}
		{\bibinfo  {journal} {Phys. Rev. Lett.}\ }\textbf {\bibinfo {volume} {107}},\
		\bibinfo {pages} {118304} (\bibinfo {year} {2011})}\BibitemShut {NoStop}%
	\bibitem [{\citenamefont {Niss}(2017)}]{nis17}%
	\BibitemOpen
	\bibfield  {author} {\bibinfo {author} {\bibfnamefont {K.}~\bibnamefont
			{Niss}},\ }\bibfield  {title} {\bibinfo {title} {Mapping isobaric aging onto
			the equilibrium phase diagram},\ }\href
	{https://doi.org/10.1103/PhysRevLett.119.115703} {\bibfield  {journal}
		{\bibinfo  {journal} {Phys. Rev. Lett.}\ }\textbf {\bibinfo {volume} {119}},\
		\bibinfo {pages} {115703} (\bibinfo {year} {2017})}\BibitemShut {NoStop}%
	\bibitem [{\citenamefont {Roed}\ \emph {et~al.}(2019)\citenamefont {Roed},
		\citenamefont {Hecksher}, \citenamefont {Dyre},\ and\ \citenamefont
		{Niss}}]{roe19}%
	\BibitemOpen
	\bibfield  {author} {\bibinfo {author} {\bibfnamefont {L.~A.}\ \bibnamefont
			{Roed}}, \bibinfo {author} {\bibfnamefont {T.}~\bibnamefont {Hecksher}},
		\bibinfo {author} {\bibfnamefont {J.~C.}\ \bibnamefont {Dyre}},\ and\
		\bibinfo {author} {\bibfnamefont {K.}~\bibnamefont {Niss}},\ }\bibfield
	{title} {\bibinfo {title} {Generalized single-parameter aging tests and their
			application to glycerol},\ }\href@noop {} {\bibfield  {journal} {\bibinfo
			{journal} {J. Chem. Phys.}\ }\textbf {\bibinfo {volume} {150}},\ \bibinfo
		{pages} {044501} (\bibinfo {year} {2019})}\BibitemShut {NoStop}%
	\bibitem [{\citenamefont {Mehri}\ \emph {et~al.}(2021)\citenamefont {Mehri},
		\citenamefont {Ingebrigtsen},\ and\ \citenamefont {Dyre}}]{meh21}%
	\BibitemOpen
	\bibfield  {author} {\bibinfo {author} {\bibfnamefont {S.}~\bibnamefont
			{Mehri}}, \bibinfo {author} {\bibfnamefont {T.~S.}\ \bibnamefont
			{Ingebrigtsen}},\ and\ \bibinfo {author} {\bibfnamefont {J.~C.}\ \bibnamefont
			{Dyre}},\ }\bibfield  {title} {\bibinfo {title} {Single-parameter aging in a
			binary {Lennard-Jones} system},\ }\href {https://doi.org/10.1063/5.0039250}
	{\bibfield  {journal} {\bibinfo  {journal} {J. Chem. Phys.}\ }\textbf
		{\bibinfo {volume} {154}},\ \bibinfo {pages} {094504} (\bibinfo {year}
		{2021})}\BibitemShut {NoStop}%
	\bibitem [{\citenamefont {B{\"o}hmer}\ \emph {et~al.}(2024)\citenamefont
		{B{\"o}hmer}, \citenamefont {Gabriel}, \citenamefont {Costigliola},
		\citenamefont {Kociok}, \citenamefont {Hecksher}, \citenamefont {Dyre},\ and\
		\citenamefont {Blochowicz}}]{boh24}%
	\BibitemOpen
	\bibfield  {author} {\bibinfo {author} {\bibfnamefont {T.}~\bibnamefont
			{B{\"o}hmer}}, \bibinfo {author} {\bibfnamefont {J.~P.}\ \bibnamefont
			{Gabriel}}, \bibinfo {author} {\bibfnamefont {L.}~\bibnamefont
			{Costigliola}}, \bibinfo {author} {\bibfnamefont {J.-N.}\ \bibnamefont
			{Kociok}}, \bibinfo {author} {\bibfnamefont {T.}~\bibnamefont {Hecksher}},
		\bibinfo {author} {\bibfnamefont {J.~C.}\ \bibnamefont {Dyre}},\ and\
		\bibinfo {author} {\bibfnamefont {T.}~\bibnamefont {Blochowicz}},\ }\bibfield
	{title} {\bibinfo {title} {Time reversibility during the ageing of
			materials},\ }\href {https://doi.org/10.1038/s41567-023-02366-z} {\bibfield
		{journal} {\bibinfo  {journal} {Nat. Phys.}\ }\textbf {\bibinfo {volume}
			{20}},\ \bibinfo {pages} {637} (\bibinfo {year} {2024})}\BibitemShut
	{NoStop}%
	\bibitem [{\citenamefont {Henot}\ \emph {et~al.}(2024)\citenamefont {Henot},
		\citenamefont {Nguyen},\ and\ \citenamefont {Ladieu}}]{hen24}%
	\BibitemOpen
	\bibfield  {author} {\bibinfo {author} {\bibfnamefont {M.}~\bibnamefont
			{Henot}}, \bibinfo {author} {\bibfnamefont {X.~A.}\ \bibnamefont {Nguyen}},\
		and\ \bibinfo {author} {\bibfnamefont {F.}~\bibnamefont {Ladieu}},\
	}\bibfield  {title} {\bibinfo {title} {Crossing the frontier of validity of
			the material time approach in the aging of a molecular glass},\ }\href
	{https://doi.org/10.1021/acs.jpclett.4c00527} {\bibfield  {journal} {\bibinfo
			{journal} {J. Phys. Chem. Lett.}\ }\textbf {\bibinfo {volume} {15}},\
		\bibinfo {pages} {3170} (\bibinfo {year} {2024})}\BibitemShut {NoStop}%
	\bibitem [{\citenamefont {Hecksher}\ and\ \citenamefont
		{Niss}(2024)}]{hecksher2024}%
	\BibitemOpen
	\bibfield  {author} {\bibinfo {author} {\bibfnamefont {T.}~\bibnamefont
			{Hecksher}}\ and\ \bibinfo {author} {\bibfnamefont {K.}~\bibnamefont
			{Niss}},\ }\bibfield  {title} {\bibinfo {title} {Single parameter aging and
			density scaling},\ }\href {https://doi.org/10.1063/5.0234620} {\bibfield
		{journal} {\bibinfo  {journal} {J. Chem. Phys.}\ }\textbf {\bibinfo {volume}
			{161}},\ \bibinfo {pages} {194504} (\bibinfo {year} {2024})}\BibitemShut
	{NoStop}%
	\bibitem [{\citenamefont {Hecksher}\ \emph {et~al.}(2019)\citenamefont
		{Hecksher}, \citenamefont {Olsen},\ and\ \citenamefont {Dyre}}]{hec19}%
	\BibitemOpen
	\bibfield  {author} {\bibinfo {author} {\bibfnamefont {T.}~\bibnamefont
			{Hecksher}}, \bibinfo {author} {\bibfnamefont {N.~B.}\ \bibnamefont
			{Olsen}},\ and\ \bibinfo {author} {\bibfnamefont {J.~C.}\ \bibnamefont
			{Dyre}},\ }\bibfield  {title} {\bibinfo {title} {Fast contribution to the
			activation energy of a glass-forming liquid},\ }\href
	{https://doi.org/10.1073/pnas.1904809116} {\bibfield  {journal} {\bibinfo
			{journal} {Proc. Natl. Acad. Sci. (USA)}\ }\textbf {\bibinfo {volume}
			{116}},\ \bibinfo {pages} {16736} (\bibinfo {year} {2019})}\BibitemShut
	{NoStop}%
	\bibitem [{\citenamefont {Nielsen}\ \emph {et~al.}(2009)\citenamefont
		{Nielsen}, \citenamefont {Christensen}, \citenamefont {Jakobsen},
		\citenamefont {Niss}, \citenamefont {Olsen}, \citenamefont {Richert},\ and\
		\citenamefont {Dyre}}]{nie09}%
	\BibitemOpen
	\bibfield  {author} {\bibinfo {author} {\bibfnamefont {A.~I.}\ \bibnamefont
			{Nielsen}}, \bibinfo {author} {\bibfnamefont {T.}~\bibnamefont
			{Christensen}}, \bibinfo {author} {\bibfnamefont {B.}~\bibnamefont
			{Jakobsen}}, \bibinfo {author} {\bibfnamefont {K.}~\bibnamefont {Niss}},
		\bibinfo {author} {\bibfnamefont {N.~B.}\ \bibnamefont {Olsen}}, \bibinfo
		{author} {\bibfnamefont {R.}~\bibnamefont {Richert}},\ and\ \bibinfo {author}
		{\bibfnamefont {J.~C.}\ \bibnamefont {Dyre}},\ }\bibfield  {title} {\bibinfo
		{title} {Prevalence of approximate $\sqrt{t}$ relaxation for the dielectric
			$\alpha$ process in viscous organic liquids},\ }\href
	{https://doi.org/http://dx.doi.org/10.1063/1.3098911} {\bibfield  {journal}
		{\bibinfo  {journal} {J. Chem. Phys.}\ }\textbf {\bibinfo {volume} {130}},\
		\bibinfo {pages} {154508} (\bibinfo {year} {2009})}\BibitemShut {NoStop}%
	\bibitem [{\citenamefont {Pabst}\ \emph {et~al.}(2021)\citenamefont {Pabst},
		\citenamefont {Gabriel}, \citenamefont {B\"ohmer}, \citenamefont {Weigl},
		\citenamefont {Helbling}, \citenamefont {Richter}, \citenamefont {Zourchang},
		\citenamefont {Walther},\ and\ \citenamefont {Blochowicz}}]{pab21}%
	\BibitemOpen
	\bibfield  {author} {\bibinfo {author} {\bibfnamefont {F.}~\bibnamefont
			{Pabst}}, \bibinfo {author} {\bibfnamefont {J.~P.}\ \bibnamefont {Gabriel}},
		\bibinfo {author} {\bibfnamefont {T.}~\bibnamefont {B\"ohmer}}, \bibinfo
		{author} {\bibfnamefont {P.}~\bibnamefont {Weigl}}, \bibinfo {author}
		{\bibfnamefont {A.}~\bibnamefont {Helbling}}, \bibinfo {author}
		{\bibfnamefont {T.}~\bibnamefont {Richter}}, \bibinfo {author} {\bibfnamefont
			{P.}~\bibnamefont {Zourchang}}, \bibinfo {author} {\bibfnamefont
			{T.}~\bibnamefont {Walther}},\ and\ \bibinfo {author} {\bibfnamefont
			{T.}~\bibnamefont {Blochowicz}},\ }\bibfield  {title} {\bibinfo {title}
		{Generic structural relaxation in supercooled liquids},\ }\href
	{https://doi.org/10.1021/acs.jpclett.1c00753} {\bibfield  {journal} {\bibinfo
			{journal} {J. Phys. Chem. Lett.}\ }\textbf {\bibinfo {volume} {12}},\
		\bibinfo {pages} {3685} (\bibinfo {year} {2021})}\BibitemShut {NoStop}%
	\bibitem [{\citenamefont {Hansen}\ \emph {et~al.}(1997)\citenamefont {Hansen},
		\citenamefont {Stickel}, \citenamefont {Berger}, \citenamefont {Richert},\
		and\ \citenamefont {Fischer}}]{hansen1997dynamics}%
	\BibitemOpen
	\bibfield  {author} {\bibinfo {author} {\bibfnamefont {C.}~\bibnamefont
			{Hansen}}, \bibinfo {author} {\bibfnamefont {F.}~\bibnamefont {Stickel}},
		\bibinfo {author} {\bibfnamefont {T.}~\bibnamefont {Berger}}, \bibinfo
		{author} {\bibfnamefont {R.}~\bibnamefont {Richert}},\ and\ \bibinfo {author}
		{\bibfnamefont {E.~W.}\ \bibnamefont {Fischer}},\ }\bibfield  {title}
	{\bibinfo {title} {Dynamics of glass-forming liquids. iii. comparing the
			dielectric $\alpha$-and $\beta$-relaxation of 1-propanol and o-terphenyl},\
	}\href@noop {} {\bibfield  {journal} {\bibinfo  {journal} {J. Chem. Phys.}\
		}\textbf {\bibinfo {volume} {107}},\ \bibinfo {pages} {1086} (\bibinfo {year}
		{1997})}\BibitemShut {NoStop}%
	\bibitem [{\citenamefont {Bauer}\ \emph {et~al.}(2013)\citenamefont {Bauer},
		\citenamefont {Burlafinger}, \citenamefont {Gainaru}, \citenamefont
		{Lunkenheimer}, \citenamefont {Hiller}, \citenamefont {Loidl},\ and\
		\citenamefont {B{\"o}hmer}}]{bauer2013debye}%
	\BibitemOpen
	\bibfield  {author} {\bibinfo {author} {\bibfnamefont {S.}~\bibnamefont
			{Bauer}}, \bibinfo {author} {\bibfnamefont {K.}~\bibnamefont {Burlafinger}},
		\bibinfo {author} {\bibfnamefont {C.}~\bibnamefont {Gainaru}}, \bibinfo
		{author} {\bibfnamefont {P.}~\bibnamefont {Lunkenheimer}}, \bibinfo {author}
		{\bibfnamefont {W.}~\bibnamefont {Hiller}}, \bibinfo {author} {\bibfnamefont
			{A.}~\bibnamefont {Loidl}},\ and\ \bibinfo {author} {\bibfnamefont
			{R.}~\bibnamefont {B{\"o}hmer}},\ }\bibfield  {title} {\bibinfo {title}
		{Debye relaxation and 250 {K} anomaly in glass forming monohydroxy
			alcohols},\ }\href@noop {} {\bibfield  {journal} {\bibinfo  {journal} {J.
				Chem. Phys.}\ }\textbf {\bibinfo {volume} {138}},\ \bibinfo {pages} {094505}
		(\bibinfo {year} {2013})}\BibitemShut {NoStop}%
	\bibitem [{\citenamefont {Gabriel}\ \emph
		{et~al.}(2018{\natexlab{a}})\citenamefont {Gabriel}, \citenamefont {Pabst},
		\citenamefont {Helbling}, \citenamefont {Böhmer},\ and\ \citenamefont
		{Blochowicz}}]{Gabriel:2018a}%
	\BibitemOpen
	\bibfield  {author} {\bibinfo {author} {\bibfnamefont {J.}~\bibnamefont
			{Gabriel}}, \bibinfo {author} {\bibfnamefont {F.}~\bibnamefont {Pabst}},
		\bibinfo {author} {\bibfnamefont {A.}~\bibnamefont {Helbling}}, \bibinfo
		{author} {\bibfnamefont {T.}~\bibnamefont {Böhmer}},\ and\ \bibinfo {author}
		{\bibfnamefont {T.}~\bibnamefont {Blochowicz}},\ }\bibfield  {title}
	{\bibinfo {title} {Nature of the {Debye}-process in monohydroxy alcohols:
			5-methyl-2-hexanol investigated by depolarized light scattering and
			dielectric spectroscopy},\ }\href
	{https://doi.org/10.1103/PhysRevLett.121.035501} {\bibfield  {journal}
		{\bibinfo  {journal} {Phys. Rev. Lett.}\ }\textbf {\bibinfo {volume} {121}},\
		\bibinfo {pages} {035501} (\bibinfo {year} {2018}{\natexlab{a}})}\BibitemShut
	{NoStop}%
	\bibitem [{\citenamefont {Gabriel}\ \emph
		{et~al.}(2018{\natexlab{b}})\citenamefont {Gabriel}, \citenamefont {Pabst},
		\citenamefont {Helbling}, \citenamefont {Böhmer},\ and\ \citenamefont
		{Blochowicz}}]{Gabriel:2018}%
	\BibitemOpen
	\bibfield  {author} {\bibinfo {author} {\bibfnamefont {J.}~\bibnamefont
			{Gabriel}}, \bibinfo {author} {\bibfnamefont {F.}~\bibnamefont {Pabst}},
		\bibinfo {author} {\bibfnamefont {A.}~\bibnamefont {Helbling}}, \bibinfo
		{author} {\bibfnamefont {T.}~\bibnamefont {Böhmer}},\ and\ \bibinfo {author}
		{\bibfnamefont {T.}~\bibnamefont {Blochowicz}},\ }\bibinfo {title}
	{Depolarized dynamic light scattering and dielectric spectroscopy: Two
		perspectives on molecular reorientation in supercooled liquids},\ in\ \href
	{https://doi.org/10.1007/978-3-319-72706-6_7} {\emph {\bibinfo {booktitle}
			{The Scaling of Relaxation Processes}}},\ \bibinfo {editor} {edited by\
		\bibinfo {editor} {\bibfnamefont {F.}~\bibnamefont {Kremer}}\ and\ \bibinfo
		{editor} {\bibfnamefont {A.}~\bibnamefont {Loidl}}}\ (\bibinfo  {publisher}
	{Springer International Publishing},\ \bibinfo {year} {2018})\ pp.\ \bibinfo
	{pages} {203--245}\BibitemShut {NoStop}%
	\bibitem [{\citenamefont {Gabriel}(2018)}]{Gabriel:2018c}%
	\BibitemOpen
	\bibfield  {author} {\bibinfo {author} {\bibfnamefont {J.}~\bibnamefont
			{Gabriel}},\ }\emph {\bibinfo {title} {Depolarisierte Dynamische
			Lichtstreuung an Monohydroxy-Alkoholen}},\ \href@noop {} {Ph.D. thesis},\
	\bibinfo  {school} {Technische Universität} (\bibinfo {year}
	{2018})\BibitemShut {NoStop}%
	\bibitem [{\citenamefont {Gabriel}\ \emph {et~al.}(2021)\citenamefont
		{Gabriel}, \citenamefont {Thoms},\ and\ \citenamefont
		{Richert}}]{gabriel2021high}%
	\BibitemOpen
	\bibfield  {author} {\bibinfo {author} {\bibfnamefont {J.~P.}\ \bibnamefont
			{Gabriel}}, \bibinfo {author} {\bibfnamefont {E.}~\bibnamefont {Thoms}},\
		and\ \bibinfo {author} {\bibfnamefont {R.}~\bibnamefont {Richert}},\
	}\bibfield  {title} {\bibinfo {title} {High electric fields elucidate the
			hydrogen-bonded structures in 1-phenyl-1-propanol},\ }\href@noop {}
	{\bibfield  {journal} {\bibinfo  {journal} {J. Mol. Liq.}\ }\textbf {\bibinfo
			{volume} {330}},\ \bibinfo {pages} {115626} (\bibinfo {year}
		{2021})}\BibitemShut {NoStop}%
	\bibitem [{\citenamefont {Böhmer}\ \emph {et~al.}(2019)\citenamefont
		{Böhmer}, \citenamefont {Gabriel}, \citenamefont {Richter}, \citenamefont
		{Pabst},\ and\ \citenamefont {Blochowicz}}]{boehmer2019influence}%
	\BibitemOpen
	\bibfield  {author} {\bibinfo {author} {\bibfnamefont {T.}~\bibnamefont
			{Böhmer}}, \bibinfo {author} {\bibfnamefont {J.~P.}\ \bibnamefont
			{Gabriel}}, \bibinfo {author} {\bibfnamefont {T.}~\bibnamefont {Richter}},
		\bibinfo {author} {\bibfnamefont {F.}~\bibnamefont {Pabst}},\ and\ \bibinfo
		{author} {\bibfnamefont {T.}~\bibnamefont {Blochowicz}},\ }\bibfield  {title}
	{\bibinfo {title} {Influence of molecular architecture on the dynamics of
			h-bonded supramolecular structures in phenyl-propanols},\ }\href@noop {}
	{\bibfield  {journal} {\bibinfo  {journal} {J. Phys. Chem. B}\ }\textbf
		{\bibinfo {volume} {123}},\ \bibinfo {pages} {10959} (\bibinfo {year}
		{2019})}\BibitemShut {NoStop}%
	\bibitem [{\citenamefont {Johari}\ and\ \citenamefont
		{Dannhauser}(1968)}]{Dannhauser1968a}%
	\BibitemOpen
	\bibfield  {author} {\bibinfo {author} {\bibfnamefont {G.~P.}\ \bibnamefont
			{Johari}}\ and\ \bibinfo {author} {\bibfnamefont {W.}~\bibnamefont
			{Dannhauser}},\ }\bibfield  {title} {\bibinfo {title} {Dielectric study of
			intermolecular association in sterically hindered octanol isomers},\ }\href
	{https://doi.org/10.1021/j100855a030} {\bibfield  {journal} {\bibinfo
			{journal} {J. Phys. Chem.}\ }\textbf {\bibinfo {volume} {72}},\ \bibinfo
		{pages} {3273} (\bibinfo {year} {1968})}\BibitemShut {NoStop}%
	\bibitem [{\citenamefont {Singh}\ and\ \citenamefont
		{Richert}(2012)}]{Singh2012}%
	\BibitemOpen
	\bibfield  {author} {\bibinfo {author} {\bibfnamefont {L.~P.}\ \bibnamefont
			{Singh}}\ and\ \bibinfo {author} {\bibfnamefont {R.}~\bibnamefont
			{Richert}},\ }\bibfield  {title} {\bibinfo {title} {Watching hydrogen-bonded
			structures in an alcohol convert from rings to chains},\ }\href@noop {}
	{\bibfield  {journal} {\bibinfo  {journal} {Phys. Rev. Lett.}\ }\textbf
		{\bibinfo {volume} {109}},\ \bibinfo {pages} {167802} (\bibinfo {year}
		{2012})}\BibitemShut {NoStop}%
	\bibitem [{\citenamefont {Sillr{\'{e}}n}\ \emph {et~al.}(2012)\citenamefont
		{Sillr{\'{e}}n}, \citenamefont {Bielecki}, \citenamefont {Mattsson},
		\citenamefont {Borjesson},\ and\ \citenamefont {Matic}}]{Sillren2012}%
	\BibitemOpen
	\bibfield  {author} {\bibinfo {author} {\bibfnamefont {P.}~\bibnamefont
			{Sillr{\'{e}}n}}, \bibinfo {author} {\bibfnamefont {J.}~\bibnamefont
			{Bielecki}}, \bibinfo {author} {\bibfnamefont {J.}~\bibnamefont {Mattsson}},
		\bibinfo {author} {\bibfnamefont {L.}~\bibnamefont {Borjesson}},\ and\
		\bibinfo {author} {\bibfnamefont {A.}~\bibnamefont {Matic}},\ }\bibfield
	{title} {\bibinfo {title} {A statistical model of hydrogen bond networks in
			liquid alcohols},\ }\bibfield  {journal} {\bibinfo  {journal} {J. Chem.
			Phys.}\ }\textbf {\bibinfo {volume} {136}},\ \href
	{https://doi.org/{10.1063/1.3690137}} {{10.1063/1.3690137}} (\bibinfo {year}
	{2012})\BibitemShut {NoStop}%
	\bibitem [{\citenamefont {Nowok}\ \emph
		{et~al.}(2021{\natexlab{a}})\citenamefont {Nowok}, \citenamefont
		{Jurkiewicz}, \citenamefont {Dulski}, \citenamefont {Hellwig}, \citenamefont
		{Ma{\l}ecki}, \citenamefont {Grzybowska}, \citenamefont {Grelska},\ and\
		\citenamefont {Pawlus}}]{nowok2021influence}%
	\BibitemOpen
	\bibfield  {author} {\bibinfo {author} {\bibfnamefont {A.}~\bibnamefont
			{Nowok}}, \bibinfo {author} {\bibfnamefont {K.}~\bibnamefont {Jurkiewicz}},
		\bibinfo {author} {\bibfnamefont {M.}~\bibnamefont {Dulski}}, \bibinfo
		{author} {\bibfnamefont {H.}~\bibnamefont {Hellwig}}, \bibinfo {author}
		{\bibfnamefont {J.~G.}\ \bibnamefont {Ma{\l}ecki}}, \bibinfo {author}
		{\bibfnamefont {K.}~\bibnamefont {Grzybowska}}, \bibinfo {author}
		{\bibfnamefont {J.}~\bibnamefont {Grelska}},\ and\ \bibinfo {author}
		{\bibfnamefont {S.}~\bibnamefont {Pawlus}},\ }\bibfield  {title} {\bibinfo
		{title} {Influence of molecular geometry on the formation, architecture and
			dynamics of h-bonded supramolecular associates in 1-phenyl alcohols},\
	}\href@noop {} {\bibfield  {journal} {\bibinfo  {journal} {J. Mol. Liq.}\
		}\textbf {\bibinfo {volume} {326}},\ \bibinfo {pages} {115349} (\bibinfo
		{year} {2021}{\natexlab{a}})}\BibitemShut {NoStop}%
	\bibitem [{\citenamefont {Nowok}\ \emph
		{et~al.}(2021{\natexlab{b}})\citenamefont {Nowok}, \citenamefont {Dulski},
		\citenamefont {Jurkiewicz}, \citenamefont {Grelska}, \citenamefont
		{Szeremeta}, \citenamefont {Grzybowska},\ and\ \citenamefont
		{Pawlus}}]{nowok2021molecular}%
	\BibitemOpen
	\bibfield  {author} {\bibinfo {author} {\bibfnamefont {A.}~\bibnamefont
			{Nowok}}, \bibinfo {author} {\bibfnamefont {M.}~\bibnamefont {Dulski}},
		\bibinfo {author} {\bibfnamefont {K.}~\bibnamefont {Jurkiewicz}}, \bibinfo
		{author} {\bibfnamefont {J.}~\bibnamefont {Grelska}}, \bibinfo {author}
		{\bibfnamefont {A.~Z.}\ \bibnamefont {Szeremeta}}, \bibinfo {author}
		{\bibfnamefont {K.}~\bibnamefont {Grzybowska}},\ and\ \bibinfo {author}
		{\bibfnamefont {S.}~\bibnamefont {Pawlus}},\ }\bibfield  {title} {\bibinfo
		{title} {Molecular stiffness and aromatic ring position--crucial structural
			factors in the self-assembly processes of phenyl alcohols},\ }\href@noop {}
	{\bibfield  {journal} {\bibinfo  {journal} {J. Mol. Liq.}\ }\textbf {\bibinfo
			{volume} {335}},\ \bibinfo {pages} {116426} (\bibinfo {year}
		{2021}{\natexlab{b}})}\BibitemShut {NoStop}%
	\bibitem [{\citenamefont {Grelska}\ \emph {et~al.}(2023)\citenamefont
		{Grelska}, \citenamefont {Jurkiewicz}, \citenamefont {Nowok},\ and\
		\citenamefont {Pawlus}}]{grelska2023computer}%
	\BibitemOpen
	\bibfield  {author} {\bibinfo {author} {\bibfnamefont {J.}~\bibnamefont
			{Grelska}}, \bibinfo {author} {\bibfnamefont {K.}~\bibnamefont {Jurkiewicz}},
		\bibinfo {author} {\bibfnamefont {A.}~\bibnamefont {Nowok}},\ and\ \bibinfo
		{author} {\bibfnamefont {S.}~\bibnamefont {Pawlus}},\ }\bibfield  {title}
	{\bibinfo {title} {Computer simulations as an effective way to distinguish
			supramolecular nanostructure in cyclic and phenyl alcohols},\ }\href@noop {}
	{\bibfield  {journal} {\bibinfo  {journal} {Physical Review E}\ }\textbf
		{\bibinfo {volume} {108}},\ \bibinfo {pages} {024603} (\bibinfo {year}
		{2023})}\BibitemShut {NoStop}%
\end{thebibliography}
\end{document}